\documentclass[a4paper,fleqn,usenatbib]{mnras}

\usepackage{newtxtext,newtxmath}

\usepackage[T1]{fontenc}
\usepackage{ae,aecompl}

\usepackage{graphicx}	
\usepackage{amsmath}	
\usepackage{amssymb}	

\usepackage{threeparttable}
\usepackage[dvipsnames,usenames,table]{xcolor}
\usepackage{multirow}

\newcommand{\sqdeg}{\mbox{deg$^{2}$}}
\newcommand{\mini}{\mbox{$M_{\rm i}$}}
\newcommand{\Mini}{\mbox{$M_{\rm i}$}}
\newcommand{\Zini}{\mbox{$Z_{\rm i}$}}

\newcommand{\Mto}{\mbox{$M_{\rm TO}$}}

\newcommand{\sub}[1]{\mbox{$_{\rm #1}$}}

\newcommand{\jks}{\mbox{$J\!-\!K_{\rm s}$}}
\newcommand{\yks}{\mbox{$Y\!-\!K_{\rm s}$}}

\newcommand{\ks}{\mbox{$K_{\rm s}$}}

\newcommand{\cmd}[3]{\mbox{$#1$ vs.\ $#2\!-\!#3$}}

\newcommand{\Mbol}{\mbox{$M_{\rm bol}$}}

\newcommand{\Msun}{\mbox{$\mathrm{M}_{\odot}$}}

\newcommand{\Mi}{\mbox{$M\sub{i}$}}
\newcommand{\Teff}{\mbox{$T_{\rm eff}$}}

\newcommand{\logg}{\mbox{$\log g$}}
\newcommand{\co}{\mbox{${\rm C/O}$}}

\newcommand{\chisq}{\mbox{$\chi^2$}}
\newcommand{\chisqlf}{\mbox{$\chi^2_{\rm LF}$}}

\newcommand{\Tbdred}{\mbox{$T_{\rm b}^{\rm dred}$}}

\newcommand{\etadust}{\mbox{$\eta_{\rm dust}$}}
\newcommand{\mdotpre}{\mbox{$\dot {M}_{\rm pre-dust}$}}
\newcommand{\mdotdust}{\mbox{$\dot {M}_{\rm dust}$}}

\newcommand{\trilegal}{\textsc{trilegal}}
\newcommand{\parsec}{\textsc{parsec}}
\newcommand{\colibri}{\textsc{colibri}}
\newcommand{\MsunYr}{\mbox{$\mathrm{M}_{\odot }\,\mathrm{yr}^{-1}$}} 

\newcommand{\WRP}{W_{\rm RP,BP}}
\newcommand{\WJK}{W_{\rm J,K}}

\defcitealias{pastorelli19}{Paper~I}

\title[TP-AGB calibration in the LMC]{Constraining the thermally pulsing asymptotic giant branch phase with resolved stellar populations in the Large Magellanic Cloud}

\author[G. Pastorelli et al.]{
Giada Pastorelli$^{1,2}$,\thanks{E-mail: gpastorelli@stsci.edu}
Paola Marigo$^{2}$,
L\'eo Girardi$^{3}$, 
Bernhard Aringer$^{2}$,
Yang Chen$^{2}$, \newauthor
Stefano Rubele$^{2,3}$, 
Michele Trabucchi$^{9,2}$, 
Sara Bladh$^{1,4}$,
Martha L.\ Boyer$^1$, \newauthor  
Alessandro Bressan$^5$,  
Julianne J.\ Dalcanton$^6$,
Martin A.T.\ Groenewegen$^7$, \newauthor  
Thomas Lebzelter$^{8}$, 
Nami Mowlavi$^{9}$,   
Katy L. Chubb$^{10}$,
Maria-Rosa L.\ Cioni$^{11}$, \newauthor 
Richard de Grijs$^{12,13,14}$,
Valentin D. Ivanov$^{15}$,
Ambra Nanni$^{16}$,
Jacco Th. van Loon$^{17}$,\newauthor
and Simone Zaggia$^{3}$
\\
$^{1}$ STScI, 3700 San Martin Drive, Baltimore, MD 21218, USA\\
$^{2}$ Dipartimento di Fisica e Astronomia Galileo Galilei, Universit\`a di Padova, Vicolo dell'Osservatorio 3, I-35122 Padova, Italy\\
$^{3}$ Osservatorio Astronomico di Padova -- INAF, Vicolo dell'Osservatorio 5, I-35122 Padova, Italy\\
$^4$ Theoretical Astrophysics, Department of Physics and Astronomy, Uppsala University, Box 516, SE-751 20 Uppsala, Sweden\\
$^5$ SISSA, via Bonomea 365, I-34136 Trieste, Italy \\
$^6$ Department of Astronomy, University of Washington, Box 351580, Seattle, WA 98195, USA \\
$^7$ Koninklijke Sterrenwacht van Belgi\"e, Ringlaan 3, 1180 Brussel, Belgium \\ 
$^8$ University of Vienna, Department of Astrophysics, Tuerkenschanzstrasse 17, A1180 Vienna, Austria \\
$^9$ Department of Astronomy, University of Geneva, Ch. des Maillettes 51, 1290 Versoix, Switzerland \\ 
$^{10}$ SRON Netherlands Institute for Space Research, Sorbonnelaan 2, 3584 CA, Utrecht, Netherlands \\
$^{11}$ Leibniz-Instit\"{u}t f\"{u}r Astrophysik Potsdam, An der Sternwarte 16, D-14482 Potsdam, Germany\\
$^{12}$ Department of Physics and Astronomy, Macquarie University, Balaclava Road, Sydney, NSW 2109, Australia \\
$^{13}$ Research Centre for Astronomy, Astrophysics and Astrophotonics, Macquarie University, Balaclava Road, Sydney, NSW 2109, Australia \\
$^{14}$ International Space Science Institute--Beijing, 1 Nanertiao, Zhongguancun, Hai Dian District, Beijing 100190, China \\
$^{15}$ European Southern Observatory, Karl-Schwarzschild-Str. 2, D-85748 Garching bei M\"{u}nchen, Germany \\
$^{16}$ Aix Marseille Universit\'e, CNRS, CNES, LAM, 38, rue Fr\'ed\'eric Joliot-Curie, F-13388 Marseille, Cedex 13 France\\
$^{17}$ Lennard-Jones Laboratories, Keele University, ST5 5BG, UK
}

\date{Accepted XXX. Received YYY; in original form ZZZ}

\pubyear{2020}

\begin{document}
\label{firstpage}
\pagerange{\pageref{firstpage}--\pageref{lastpage}}
\maketitle

\begin{abstract}
Reliable models of the thermally pulsing asymptotic giant branch (TP-AGB) phase are of critical importance across astrophysics, including our interpretation of the spectral energy distribution of galaxies, cosmic dust production, and enrichment of the interstellar medium.
With the aim of improving sets of stellar isochrones that include a detailed description of the TP-AGB phase, we extend our recent calibration of the AGB population in the Small Magellanic Cloud (SMC) to the more metal rich Large Magellanic Cloud (LMC).
We model the LMC stellar populations with the \trilegal\ code, using the spatially-resolved star formation history derived from the VISTA survey.
We characterize the efficiency of the third dredge-up by matching the star counts and the \ks-band luminosity functions of the AGB stars identified in the LMC.
In line with previous findings, we confirm that, compared to the SMC,  the third dredge-up in AGB stars of the LMC is somewhat less efficient, as a consequence of the higher metallicity. The predicted range of initial mass of C-rich stars is between $\Mini \approx 1.7 - 3~\Msun$ at $\Zini = 0.008$.
We show how the inclusion of new opacity data in the carbon star spectra will improve the performance of our models. We discuss the predicted lifetimes, integrated luminosities and mass-loss rate distributions of the calibrated models.
The results of our calibration are included in updated stellar isochrones publicly available.
\end{abstract}




\section{Introduction}
\label{intro}

Close to the end of their lives, low- and intermediate-mass stars, with initial masses between approximately 0.8~\Msun up to 6-8~\Msun,
evolve through the thermally pulsing asymptotic giant branch (TP-AGB) phase \citep{Herwig05}. Despite the very short duration of this evolutionary phase (less than a few Myr), TP-AGB stars can contribute significantly to the integrated luminosity of intermediate-age stellar populations, and the treatment of the TP-AGB phase can affect the interpretation of the spectral energy distribution of 
unresolved galaxies, in particular the derivation of their stellar mass and age \citep{Maraston_etal_06, Conroy13, Zibetti_etal_13, villaume15}. 
Furthermore, TP-AGB stars might be significant dust producers in the local Universe and at high redshift \citep[for extensive discussions see e.g.][]{boyer12, nanni18, schneider14, SR16, valiante09, zhukovska08, zhukovska13}, and can be important contributors to the chemical enrichment of galaxies \citep{kobayashi11, karakas_lattanzio14}.

TP-AGB stars are also useful to probe the star formation history in other galaxies when deep photometry is not available or not possible, as shown by \citet{javadi13,javadi17, rezaeikh14, HamedaniGolshan17,Hashemi19} for M33 and other Local Group galaxies. TP-AGB stars might also offer an additional way to improve the calibration of the extragalactic distance scale either using their long period variability \citep{pierce00,huang18,huang20} or the mean photometric properties of the carbon stars \citep{ripoche20,madore20}.

Despite their widespread importance, TP-AGB models suffer from large uncertainties and present models from various authors differ significantly in many important outcomes, such as stellar lifetimes, initial mass of carbon stars, and chemical yields.
The main sources of uncertainties can be traced to i) the lack of a robust theory of convection, which affects mixing processes such as third dredge-up (3DU) and hot bottom burning (HBB), and ii) the difficulty of modelling the physics of stellar winds, and therefore the mass-loss rate as a function of the stellar parameters, which control the TP-AGB lifetimes \citep{Marigo_15}.

In recent years significant progress has been made on both theoretical and observational sides. Now we have the possibility to exploit complete samples of resolved AGB stars from the optical to the infrared wavelengths, observed in stellar systems that span a wide range of metallicities, from the metal poor dwarf galaxies \citep{dalcanton09, dalcanton12, boyer15_dustingsI, boyer17} to the metal rich M31 \citep[and Goldman et al., in prep.]{boyer19}, and for which we have robust measurements of the star formation history (SFH) \citep[e.g.][]{weisz14, lewis15, williams17}.

One fundamental laboratory to study AGB star populations can be found in the Small and Large Magellanic Clouds (SMC and LMC). The stellar populations of these two  
irregular galaxies are very well studied, and thanks to the numerous spectroscopic and photometric surveys carried out in recent years, we have a complete sample of AGB stars for which we also have a reliable identification of their chemical type, i.e. carbon-rich (C-rich) and oxygen-rich (O-rich). 

On the theoretical side, so-called `full stellar models' are calculated by integrating the stellar structure equations across the whole star, hence resolving the physical structure from the centre to the surface.  
Moreover, full models still need to rely on parametrized descriptions of complicated three-dimensional processes like convection, overshoot, mass loss.  Predictions for the TP-AGB, in particular,  are significantly affected by numerical details that may differ from author to author \citep[e.g. this is the case of the third dredge-up;][]{FrostLattanzio_96}.
Furthermore, the calculation of full models is time consuming, which makes it difficult to efficiently explore and test the wide range of parameters necessary to provide a thorough calibration of the uncertain processes. 

In this context, a complementary approach is provided by the so-called `envelope models', for which the TP-AGB evolution is calculated by including analytical prescriptions (derived from full model calculations), complemented with envelope integrations. 
In this work we use the \colibri\ code, fully described in \citet{marigo13}. \colibri\ combines a synthetic module that includes the free parameters to be calibrated with the aid of observations, i.e. mass loss, and 3DU occurrence and efficiency, coupled with a complete envelope integration of the stellar structure equations from the atmosphere down to the bottom of the hydrogen-burning shell. 
This allows the code to follow the changes in the envelope and atmosphere structures (e.g. driven by chemical composition changes) with the same level of detail as in full models, but with a computational time that is typically two orders of magnitude shorter \citep[see e.g. figure 10 of][]{marigo13}. This feature is fundamental to efficiently explore the range of parameters that need to be calibrated as a function of stellar mass and metallicity. 

By combining the computational agility of the \colibri\ code, and the detailed stellar population synthesis simulations produced with the \trilegal\ code \citep{girardi05}, we can test different mass-loss prescriptions and put quantitative constraints on the occurrence and efficiency of the 3DU. This is achieved by reproducing the star counts and the luminosity functions (LFs) of an observed sample of AGB stars with known SFH. The approach was pioneered by \citet{groenewegen1993}, \citet{marigo99} and \citet{marigo07}, and more recently adopted by \citet{girardi10} and \citet{rosenfield14} using AGB samples in dwarf galaxies from the ANGST survey. 
In a recent paper, \citet[][hereafter Paper~I]{pastorelli19} applied the same approach to the population of AGB stars in the SMC classified by \citet[][hereafter B11]{boyer11} and \citet{SR16}. The initial metallicity range covered by such work is below $\Zini = 0.008$.

In this work, we extend the calibration of our TP-AGB models to higher metallicities using the observed sample of AGB stars classified by B11 from the \textit{Spitzer} program `Surveying the Agents of a Galaxy's Evolution in the LMC' \citep[SAGE-LMC,][]{blum06, meixner06}. Starting from the best-fitting model for the SMC, we compute additional evolutionary tracks with initial metallicity $\Zini \geq 0.008$, with the aim of reproducing at the same time the star counts and the luminosity functions of the whole TP-AGB population, and the C- and O-rich samples. 

The paper is organized as follows. We briefly recall the general scheme of our calibration strategy, and we describe the input data and AGB observations in Sect.~\ref{sec:data}. The adopted stellar models are presented in Sect.~\ref{sec:models}. We present the results of our LMC calibration in Sect.~\ref{sec:results} and we discuss them in Sect.~\ref{sec:discussion}. Final remarks close the paper in Sect.~\ref{sec:conclu}.

\section{Data and methods}
\label{sec:data}

As thoroughly described in \citet{pastorelli19} in the case of the SMC galaxy, our work relies on three fundamental components, now regarding the LMC galaxy: 

\begin{enumerate} 
\item \label{item:sfh} The spatially-resolved SFH derived 
for well-defined subregions of the sky across the LMC. Importantly, the SFH is derived from regions of the colour-magnitude diagram (CMD) in which the expected number of TP-AGB stars is negligible.
This way we prevent the SFHs to be affected by uncertainties in the TP-AGB models themselves.
Moreover, the same SFH-recovery process produces estimates of the distance and extinction of each subregion.
\item \label{item:cat} AGB catalogues that include an accurate identification of C- and O-rich type stars, mainly derived from a combination of Two Micron All-Sky Survey (2MASS) and \textit{Spitzer} photometric data, and complemented by additional information, e.g. from spectroscopic surveys.
\end{enumerate}
Regions of the LMC with available data for both the items above are modelled with the \trilegal\ stellar population synthesis code. This procedure makes use of the SFH, distances, and extinctions from item \ref{item:sfh}, and produces the theoretical counterparts of the catalogues from item \ref{item:cat}.
Such simulations make use of our third basic component: 
\begin{enumerate}
\setcounter{enumi}{2}
\item \label{item:models} Extended grids of evolutionary models for TP-AGB stars, computed for different choices of the parameters describing to 3DU events and mass loss. 
\end{enumerate}
In the following, we briefly describe components \ref{item:sfh} and \ref{item:cat}. The models from \ref{item:models} will be introduced one-by-one in Sects.~\ref{sec:models} and \ref{sec:results}, together with their comparison with the catalogues from \ref{item:cat}.

\subsection{Star Formation History}
\label{ssec:sfh}
\begin{figure}
    \centering
    \includegraphics[width=\columnwidth]{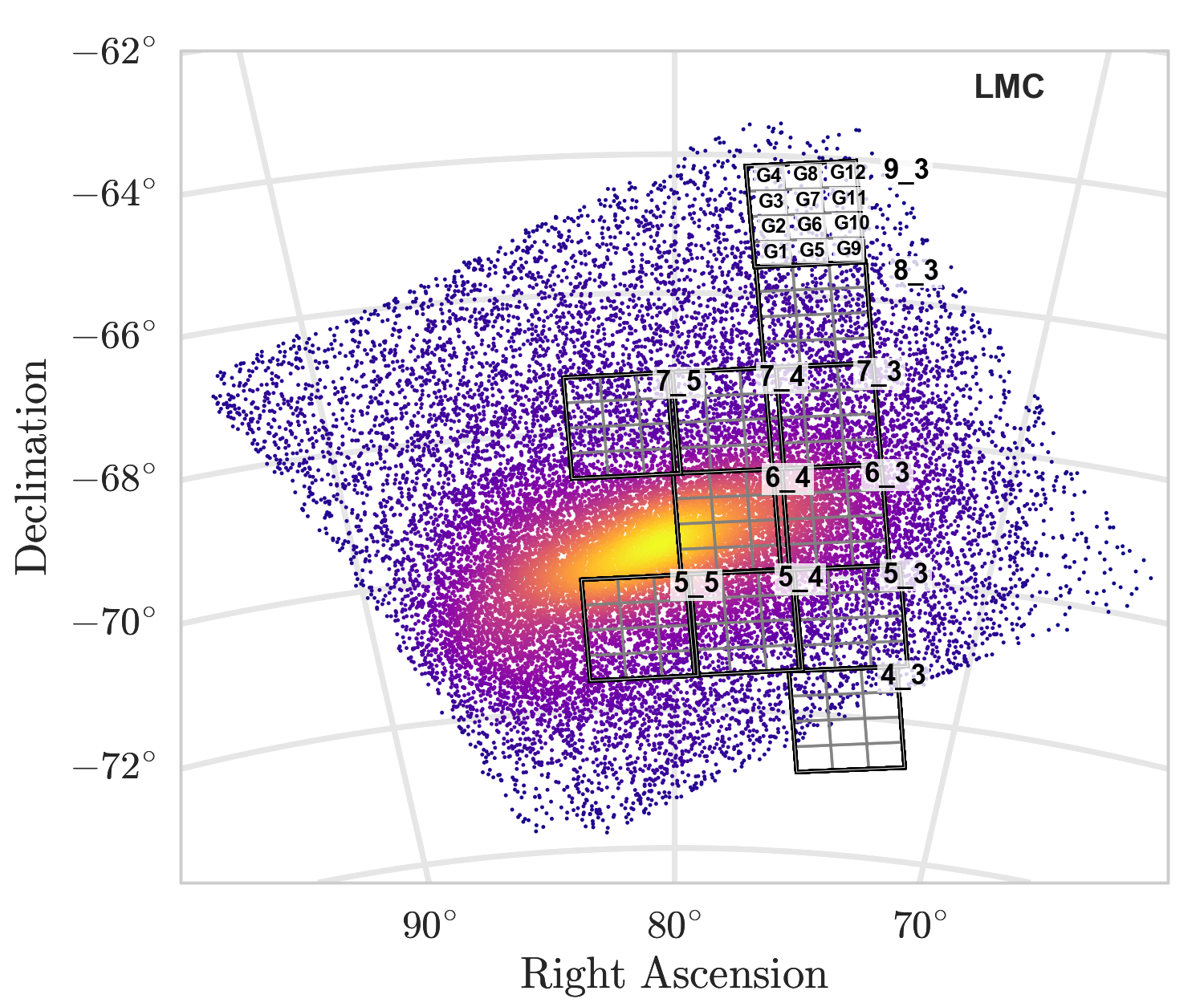}
    \caption{VMC tiles for which the SFH recovery is available. Each tile (in black) is subdivided into 12 subregions (in grey), as illustrated for the tile LMC 9\_3. The background image shows the density map of the AGB stars classified by B11.}
    \label{fig:vmc_sage_map}
\end{figure}

We use the SFH derived from deep near-infrared data ($J$, $Y$, \ks\ filters) from the VISTA survey of the Magellanic Clouds \citep[VMC;][]{cioni11}. We refer to \citet[][and references therein]{rubele18} for a complete description of the method, and its application to VMC data for the SMC. Briefly, the SFH is derived from two different CMDs, \cmd{\ks}{J}{\ks} and \cmd{\ks}{Y}{\ks}, by finding the model that minimizes a data--model \chisq-like statistic. The model CMDs are built with the \trilegal\ population synthesis code (the version by \citealt{marigo17}), using \parsec\ v1.2S \citep{bressan12} stellar evolutionary tracks. The method fits not only the stellar mass formed in several age bins, but also the age-metallicity relation (AMR), distance, and mean extinction for each analysed subregion. Throughout this work, `SFH' refers to the combination of the star formation rate and metallicity as a function of age. The derivation of the SFH of the LMC will be fully described in a forthcoming paper.

Fig.~\ref{fig:vmc_sage_map} shows the VMC tiles for which the SFH is currently available in the LMC, superimposed on the density map of the AGB stars classified by B11. Table~\ref{tab:tab_tiles} summarizes their central coordinates, and the number of AGB stars in each tile. These tiles are about $1.5^\circ\times1.0^\circ$ large\footnote{The size of the tiles corresponds to the area where each pixel of the tile image is observed at least twice.}, and their longer dimension runs almost along the North-South direction \citep{cioni11}. For the SFH analysis, each tile is divided into 12 subregions, labelled from G1 to G12 as indicated in Fig.~\ref{fig:vmc_sage_map}.

\begin{table}
\centering                         
\caption{Central sky coordinates of the VMC tiles and number of AGB stars identified by B11 for each tile. }
\label{tab:tab_tiles}
\begin{tabular}{cccc}
\hline
 Tile$^1$  & R.A.$_{\mathrm {J2000}}$ & Dec.$_{\mathrm {J2000}}$ & N. AGB \\
 & (h:m:s)  & (d:m:s)  &  \\ 
 \hline
LMC 4\_3 &   04:55:19.5   & $-$72:01:53.4    & 130  \\
LMC 5\_3 &   04:56:52.5   & $-$70:34:25.7    & 628  \\
LMC 5\_4 &   05:10:41.5   & $-$70:43:05.9    & 1070 \\
LMC 5\_5 &   05:24:30.3   & $-$70:48:34.2    & 1558 \\
LMC 6\_3 &   05:00:42.2   & $-$69:08:54.2    & 1634 \\
LMC 6\_4 &   05:12:55.8   & $-$69:16:39.4    & 2852 \\
LMC 7\_3 &   05:02:55.2   & $-$67:42:14.8    & 920  \\
LMC 7\_4 &   05:14:06.4   & $-$67:49:21.7    & 851  \\
LMC 7\_5 &   05:25:58.4   & $-$67:53:42.0    & 647  \\
LMC 8\_3 &   05:04:55.0   & $-$66:15:29.9    & 458  \\
LMC 9\_3 &   05:06:40.6   & $-$64:48:40.3    & 155  \\
\hline
&\multicolumn{2}{c}{Total area}   &  \multicolumn{1}{c}{N$_{\mathrm{Tot}} $  AGB}   \\
&\multicolumn{2}{c}{$\approx 25\,\sqdeg $}    & \multicolumn{1}{c}{10903}   \\
\hline
\multicolumn{4}{l}{ {\bf Notes:} } \\
\multicolumn{4}{l}{$^{(1)}$  Excluded subregions: \mbox{LMC 4\_3} G1, G2, G5, G6, G9, G10, G11}\\
\end{tabular}
\end{table}

\begin{figure*}
    \centering
    \includegraphics[width=0.7\textwidth]{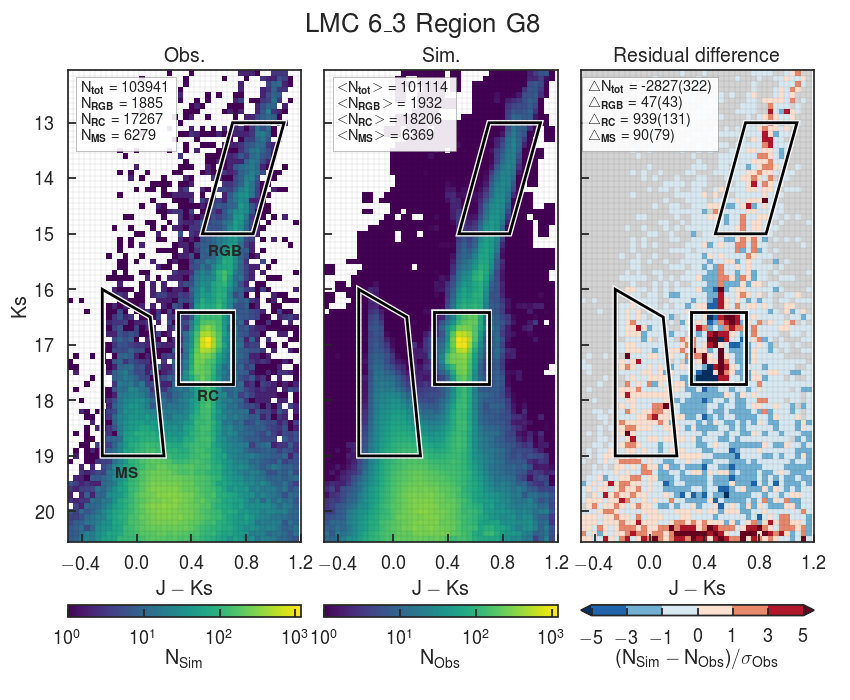}
    \caption{Comparison between the observed VMC data (left panel) and the simulated Hess diagram (middle panel) derived from the \ks\ vs. \jks\ solution for the subregion G8 of tile LMC 6\_3.
    The simulated diagram is the median of 10 \trilegal\ realizations. 
    The right panel shows the difference between the model and the observed data, divided by the square root of the observed number counts ($\sigma_{\mathrm{Obs}}$), so as to reveal the CMD regions with the most significant residuals. The boxes correspond to the RGB, Red Clump (RC) and upper Main Sequence (MS) regions. The total number of stars, and the number of stars in each box, are reported in the legend of each panel. In addition, the legend in the right panel indicates, in brackets, the $1~\sigma$ value of the observed number counts in the corresponding box. }
    \label{fig:hess6308}
\end{figure*}

\begin{figure*}
    \centering
    \begin{minipage}{0.49\textwidth}
    \includegraphics[width=0.9\columnwidth]{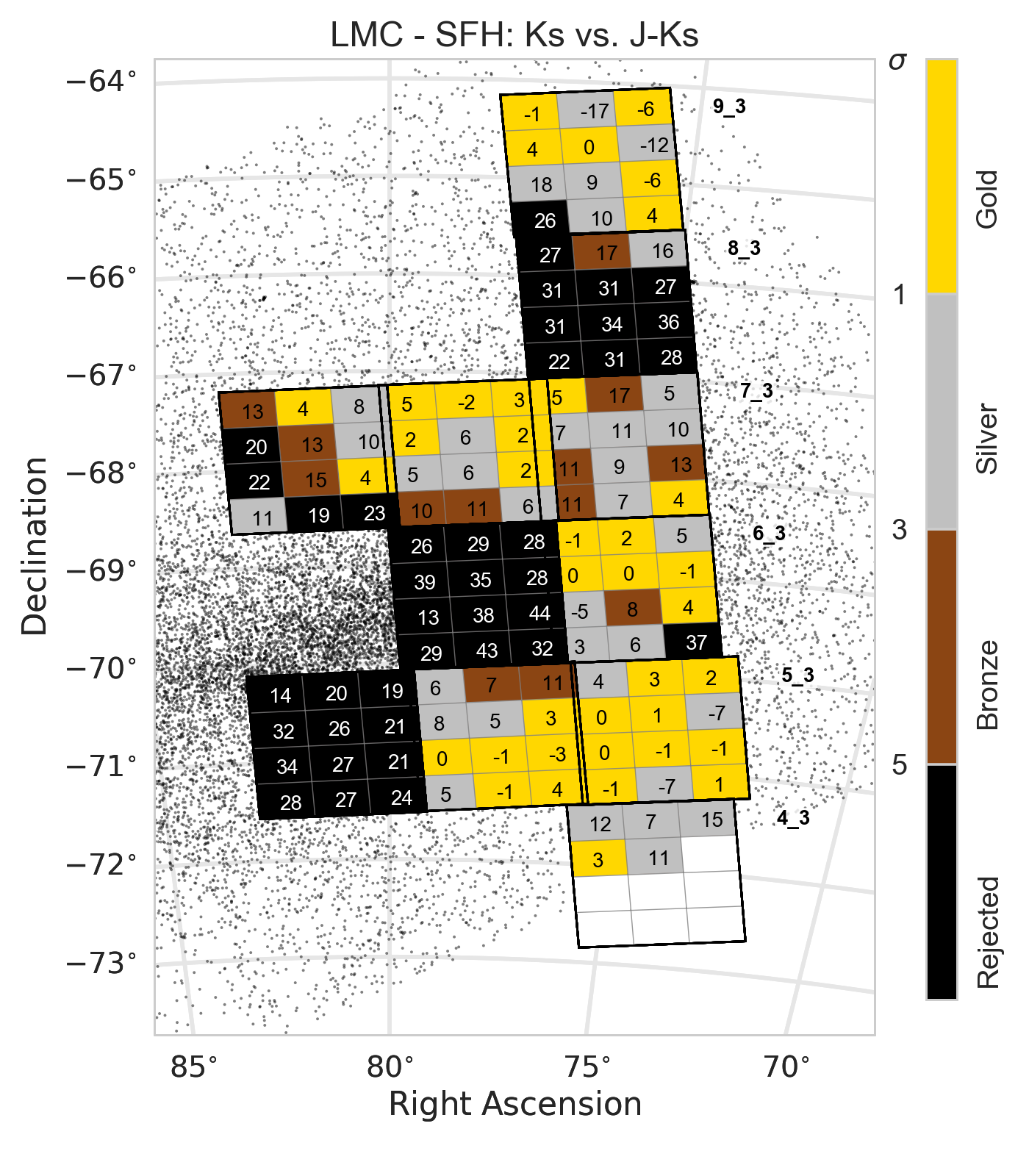}
    \end{minipage}%
    \begin{minipage}{0.49\textwidth}
    \includegraphics[width=0.9\columnwidth]{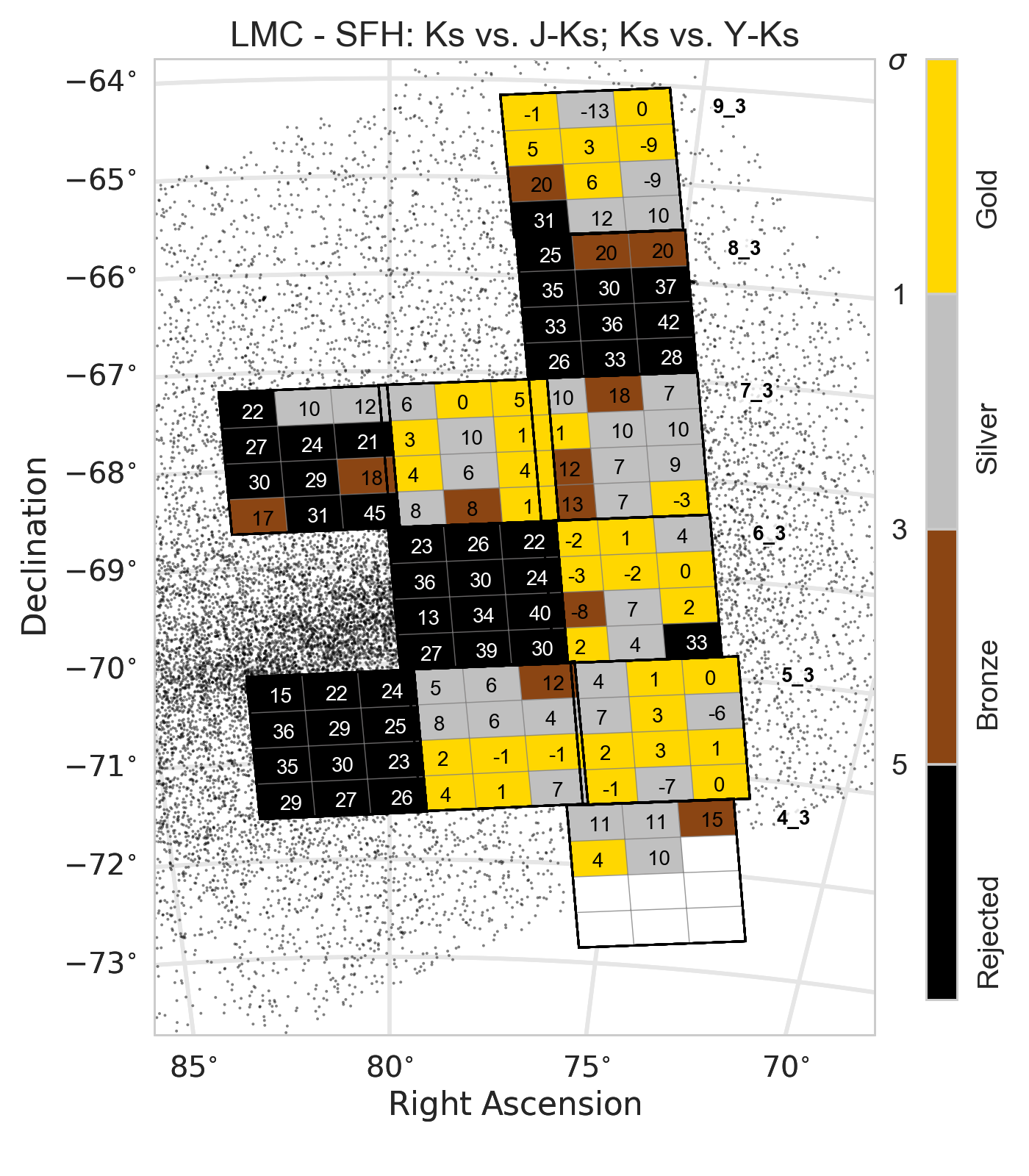}
    \end{minipage}
      \caption{Maps of the predicted vs. observed RGB star counts across the LMC area considered in this work. The two maps show the results obtained with the input SFH solutions coming from the \ks\ vs. \jks\ CMDs (left panel), and from the combined use of \ks\ vs. \jks\ and \ks\ versus \yks\ CMDs (right panel). For each subregion we report the  percentage of the difference between the number of simulated and observed RGB stars, i.e.\ $100\times (N^{\mathrm{RGB}}_\mathrm{sim} - N^{\mathrm{RGB}}_\mathrm{obs}) / N^{\mathrm{RGB}}_\mathrm{obs} $. 
      The color-code of each subregion is the following:
       `Gold' for RGB star counts within $1\sigma$, `Silver' for star counts between 1 and $3\sigma$, `Bronze' for star counts between 3 and $5\sigma$. Subregions with differences in RGB star counts above $5\sigma$ are not considered in the calibration.}   
      \label{fig:rgb_good}
\end{figure*}

In the case of the LMC, the observed photometry can be significantly affected by crowding, which in turn affects the SFH robustness. 
For this reason, we use the SFH solutions derived from the \ks\ vs. \jks\ and \ks\ vs. \yks\ CMDs to simulate the VMC data of the LMC for each subregion to assess the quality of the solutions. For each subregion we compute 10 \trilegal\ simulations and we compare the median number of simulated red giant branch (RGB) stars with the observed one in the \ks\ vs. \jks\ CMD, as shown in Fig.~\ref{fig:hess6308}. 
The CMD region used to select RGB stars is such that 1) we avoid AGB contamination, 2) we exclude CMD regions that are likely to be severely affected by crowding errors and incompleteness.
The results of these tests are shown in Fig.~\ref{fig:rgb_good}, plotted in terms of a fractional error in the RGB counts and standard deviations from the expected numbers.

From these figures, it is evident that some subregions (and even entire tiles) present large errors in their predicted RGB star counts. In some tiles (in particular the LMC 5\_5, 6\_4 and 7\_5), the discrepancies are likely linked to the severe crowding conditions close to the LMC Bar, which has probably affected the SFH derivation in unexpected ways.
It is beyond the scope of this paper to delve into the possible causes of these discrepancies. 
Moreover, there would be no easy solution to these problems given the large amount of computer time involved in performing the PSF photometry, and millions of artificial star tests over the VMC images, necessary to perform the SFH-recovery \citep[see][]{rubele18}. 
Based on these tests, we culled our list of LMC subregions, keeping those for which the errors in RGB star counts are smaller than a given threshold. To this aim, we classify our subregions in four broad categories: `Gold' are those in which RGB star counts are reproduced within 1$\sigma$, `Silver' are between 1$\sigma$ and $3\sigma$, and 
`Bronze' are between 3$\sigma$ and $5\sigma$, as illustrated in Fig.~\ref{fig:rgb_good}. Subregions with RGB star counts errors above 5$\sigma$ (`Rejected' regions) are not considered in the calibration.
We notice that the mismatch between model and data RGB counts is generally smaller or similar to $\approx\!10$ per cent for Gold and Silver subregions.

\subsection{Selected areas and their AGB numbers}
\label{ssec:sel_AGB}

Table~\ref{tab:sfh_check_summary} shows the number of AGB stars that we can use in the calibration, depending on whether we choose to use 1) Gold, 2) Gold + Silver, or 3) Gold + Silver + Bronze subregions. We also distinguish between whether the SFH solution is obtained from the \ks\ vs. \jks\ CMD alone, or from both the \ks\ vs. \jks\ and the \ks\ vs. \yks\ CMDs.

There is obviously a trade-off between adopting more inclusive criteria, and including regions with larger errors in their SFHs (as indicated by the mismatches in their RGB star counts). 
In this work, we consider the Gold+Silver VMCs subregions, for which the RGB star counts are reproduced within $3\sigma$.
Moreover, we decide to use the SFH solutions obtained from the \ks\ vs. \jks\ CMD alone, so to maximize the number of AGB stars available, i.e.\ 4664 sources in 72 subregions. 
Despite the total number of AGB stars is lower than the number used in our previous work for the SMC \citep{pastorelli19}, it is large enough to reach a TP-AGB calibration of a similar quality, even with the present partial coverage of the LMC galaxy. 
 
\begin{table}
    \centering
   \caption{Number of VMC subregions for which the simulated RGB star counts agree within 1$\sigma$ ('Gold'), 3$\sigma$ ('Gold+Silver'), and 5$\sigma$ ('Gold+Silver+Bronze') of the observed RGB star counts. The resulting number of AGB stars from the B11 catalogues are shown. The results are presented for the SFH solutions from the \jks\ and \jks;\yks\ CMDs.}    
    \begin{tabular}{clcc}
    \hline
    & \multicolumn{3}{c}{CMD} \\
                 & &\ks\ vs. \jks   &   \ks\ vs. \jks ; \ks\ vs. \yks \\ 
\hline     
\multicolumn{2}{c}{N$_{\rm reg.}$ tot. }     &   125  &   125  \\
\multirow{2}{*}{$1 \sigma$}& N$_{\rm reg.}$  &  38  &    36  \\ 
&N$_{\rm AGB}$ &  2404  &  2478  \\
\multirow{2}{*}{$3 \sigma$}& N$_{\rm reg.}$ &    72  &    69  \\
& N$_{\rm AGB}$    &  4664  &  4648   \\
\multirow{2}{*}{$5 \sigma$}& N$_{\rm reg.}$ &   85  &    81  \\
& N$_{\rm AGB}$    &  5742  &  5555   \\
\hline
    \end{tabular}
    \label{tab:sfh_check_summary}
\end{table}


\subsection{Observations of AGB stars in the LMC}
\label{sec:LMCdata}

The calibration performed in this work is based on the LMC AGB population classified by  B11. They combined data from the 2MASS \citep{skrutskie06} and SAGE-LMC surveys to study the evolved population of the LMC, and to give a photometric classification of the AGB stars. 
The area covered by the SAGE-LMC survey is shown in Fig.~\ref{fig:vmc_sage_map}.

The SAGE-LMC catalogue is a complete census of AGB stars in the LMC, including optically visible O- and C-rich stars above the tip of the RGB, as well as the most obscured dusty sources. 
The AGB stars are classified in O-rich (or O-AGB), C-rich (or C-AGB), extreme-AGB (X-AGB), and anomalous-AGB (a-AGB).
 
O- and C-rich sources are classified based on their position in the \cmd{\ks}{J}{\ks} CMD.   
The class of X-AGBs, first introduced by \cite{blum06}, includes the very dusty stars, empirically selected on the basis of their $J-[3.6]$ and $[3.6] - [8.0]$ colours. Most of them are C-rich stars, but a small number of O-rich is also present \citep{vanloon05}.

The chemical type of sources classified as a-AGB cannot be photometrically inferred from the available combinations of 2MASS and \textit{Spitzer} colours. However, for a subsample of them, \citet[][hereafter B15]{boyer15} used optical spectra from \cite{olsen11} to determine their spectral type in both SMC and LMC. 
In the LMC, the percentage of a-AGB stars that are O-rich is about 77 per cent, whereas in the SMC the percentage is about 50 per cent. 
The total number of a-AGB stars spectroscopically analysed by B15 in the LMC is 613, leaving about 5800 a-AGB stars not classified as O- or C-rich. On the basis of the selection criteria adopted in Sect.~\ref{ssec:sel_AGB}, the number of a-AGBs considered here is 1076.  

\begin{figure*}
    \centering
    \includegraphics[width=0.9\textwidth]{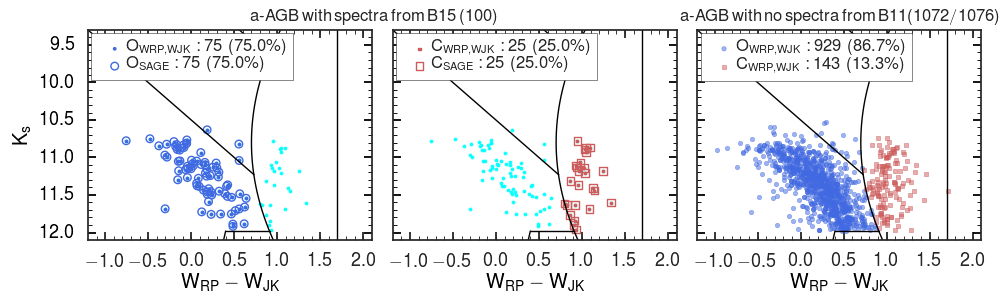}
    \caption{Location of the a-AGB stars in the \textit{Gaia}-2MASS diagram. The black lines correspond to the regions described by \citet{Lebzelter_etal18} with the bending curve dividing O-rich (left) and C-rich (right) stars. 
    The sample of a-AGBs with spectra from B15 are shown as cyan dots in the left and middle panels.
    Left panel: a-AGB stars classified as O-rich by B15 are marked with blue open circles (O$_{\rm SAGE}$), and those classified as O-rich in the \textit{Gaia}-2MASS diagram (O$_{\rm WRP,WJK}$) as blue dots. Middle panel: a-AGB stars classified as C-rich by B15 are shown as red open squares (C$_{\rm SAGE}$), and those classified as C-rich in the \textit{Gaia}-2MASS diagram (C$_{\rm WRP,WJK}$) as red squares. Right panel: a-AGB stars, for which spectroscopic information is not available, classified using the \textit{Gaia}-2MASS diagram. The legend in each panel shows the number of classified sources and their percentage with respect to the total. Only the sources selected for this work are shown. The number of a-AGB stars with spectra from B15 is 100, all of them have a \textit{Gaia} DR2 counterpart. The number of a-AGBs from B11 with no spectra is 1076, and we find a \textit{Gaia} DR2 match for 1072 of them.}
    \label{fig:aAGB_WRPWJK_class}
\end{figure*}

In \citetalias{pastorelli19}, we took into account the contribution of a-AGB stars by weighting the observed luminosity functions of the C-AGB and O-AGB using the fraction of a-AGB sources spectroscopically classified as either C-rich or O-rich. 
Here, we apply a diagnostic tool proposed by \citet{Lebzelter_etal18} to assign the chemical type to the sample of a-AGBs. We refer to this diagram as \textit{Gaia}-2MASS diagram, as it combines the \textit{Gaia} Data Release 2 \citep[DR2;][]{gaiaDR2} and 2MASS Wesenheit functions\footnote{The expressions for the \textit{Gaia} and 2MASS Wesenheit functions are

 $\WRP= G_{\mathrm{RP}} -1.3\times(G_{\mathrm {BP}} - G_{\mathrm{RP}})$ and $\WJK = \ks - 0.686\times(J-\ks)$ respectively.}.

To construct the \textit{Gaia}-2MASS diagram for the a-AGB sample, we cross-match the SAGE-LMC catalogue with the \textit{Gaia} DR2 data. 
For each source with both J and \ks-band magnitudes from 2MASS, we obtain the \textit{Gaia} counterpart using a search radius of 5$\arcsec$. We further check our results using the pre-computed cross-match of \textit{Gaia} DR2 with 2MASS \citep[`2MASS BestNeighbour' table,][]{gaiaXmatch}.
We find 1072 matches out of 1076 a-AGBs. 

We first compare the classification from the \textit{Gaia}-2MASS diagram with the results of B15 for the a-AGBs with a spectroscopic classification (see left and middle panels of Fig.~\ref{fig:aAGB_WRPWJK_class}). The \textit{Gaia}-2MASS method is in perfect agreement with the spectroscopic one for the sample of a-AGBs considered, with all the available sources correctly classified.
Given these results, we classify the a-AGB stars for which no spectral information is available from B15 according to the position in the \textit{Gaia}-2MASS diagram as shown in the right panel of Fig.~\ref{fig:aAGB_WRPWJK_class}. 

We perform a further check on all the sources for which the classification is only based on the photometry from B11 and from the \textit{Gaia}-2MASS diagram. 
We use the spectroscopic classification from \citet{groenewegen_sloan_18} based on \textit{Spitzer} IRS spectra, the C-star catalogue by \citet{kontizas01}, and the catalogue of MK spectral types compiled by \citet{MKtypes}.
By using a search radius of 2$\arcsec$, we find a total of 807 counterparts. 
We correct the photometric classification of 3 C-rich and 47 O-rich stars that are spectroscopically classified as M- and C-type, respectively. 

Table~\ref{tab:sage_stat} lists the final number counts of C-, O-, X-, and a-AGB stars used in this work. 
The contribution of the remaining 4 a-AGBs with no counterpart in \textit{Gaia} DR2 is taken into account by weighting the LFs as in \citetalias{pastorelli19}, using the fraction of O- and C-rich a-AGBs from B15.

\begin{table}
\centering
\caption{Number counts of C-, O-, X-, and a-AGB stars. The final number counts used in this work include the contribution of a-AGB stars from B15 without \textit{Gaia} DR2 counterparts.}
\label{tab:sage_stat}
\begin{tabular}{lcc}
\hline
Population &  N. star & N. star (this work)\\
\hline
C-AGB        & 1453 &  1454 \\  
O-AGB        & 2931 &  2934  \\ 
X-AGB        & 276  &   276\\ 
a-AGB        & 4    &    0 \\  
\hline
\end{tabular}
\end{table}

\subsection{TRILEGAL simulations and model selection criteria}
\label{ssec:data_model_comp}
 
Our calibration strategy, including the details of the AGB selection criteria adopted in the models are extensively described in \citetalias{pastorelli19}. We briefly summarize them here.

We simulate the photometry of each LMC subregion selected in Sect.~\ref{ssec:sel_AGB} with the \trilegal\ code, and we merge all the synthetic catalogues to be compared with the observed AGB catalogue.
This latter only includes the sources located in the same sky area as the VMC subregion.

Each subregion is modelled according to its SFR, AMR, distance, and reddening derived from the SFH recovery. 
We adopt the \citet{kroupa01} initial mass function for single stars, and we simulate non-interacting binaries using a binary fraction of 30 per cent along with a uniform mass distribution of mass ratios between the secondary and the primary components in the range 0.7 - 1.
The photometric errors are taken into account in \trilegal\ following the distribution of errors as a function of magnitude reported in the observed catalogue. 

The Milky Way foreground and incompleteness of the data are not simulated, simply because these effects are less of a problem in the CMD region occupied by AGB stars.
The 2MASS catalogue is complete down to $\ks = 14.3$~mag \citep{skrutskie06}, that is $\approx 2$~mag fainter than the tip of the RGB in the LMC. Furthermore, AGB stars are identified by combining both near- and mid-infrared photometry (B11). This ensures that the most obscured dusty stars are not missed from the observed catalogues.  \citet{melbourne13} estimated the foreground contamination for the AGB population in the LMC to be below 1 per cent.

To select AGB stars and the three classes of C-, O-, and X-AGB in the synthetic catalogues we use a combination of theoretical parameters and photometric criteria. The C/O ratio is used to select C- and O-rich stars. The class of O-rich stars contains both early-AGB and TP-AGB stars and we use the same photometric criteria as in B11. X-AGB stars are selected using photometric criteria alone. We refer to Section 2.5 and Appendix A.2 in \citetalias{pastorelli19} for a complete description. 

We produce 10 \trilegal\ realizations for each subregion, and we calculate the \chisq\ of the median LF, \chisqlf, with respect to the data.  
A satisfactory agreement between data and model is achieved when we obtain the lowest \chisqlf\ values for the entire sample, and for the three classes of AGB stars. 

In \citetalias{pastorelli19}, we identified a systematic shift in the \jks\ colour of the synthetic RSG and O-rich stars in the SMC.  The most likely explanation for this discrepancy is a temperature offset. A mismatch between the predicted and observed colours may hamper our comparison with the observed O-rich stars, specifically the RSG and O-rich separation, which is based on photometric criteria. 
\citetalias{pastorelli19} corrected the synthetic photometry of RSGs and O-rich AGBs to match the observed RSG colours. In this work we avoid making such corrections as in the case of the LMC we do not not find a significant shift in the RSG colour. However,  the O-rich AGBs show a similar shift to redder colours comparable to the SMC case, i.e. $\Delta \jks \approx 0.05 - 0.1$ between $\ks \approx 12.7 - 11.8$~mag.
While an assessment of this discrepancy will be the subject of a future work, we emphasize that this only affects the number of O-rich stars by about 5 per cent. 

\section{Stellar evolutionary models}
\label{sec:models}

We adopt the \parsec\ database \citep{bressan12} of stellar evolutionary models to cover all phases from the pre-main sequence up to carbon ignition in massive stars, or up to the occurrence of the first thermal pulse in low- and intermediate-mass stars. For these latter stars the TP-AGB phase is then computed with the \colibri\ code \citep{marigo13}.
In this section, we recall the main features of our TP-AGB models, and we refer to \citet{marigo13} and \citet{pastorelli19} for a detailed description. 

The \colibri\ calculations start from the stellar configuration given by the \parsec\ models at the beginning of TP-AGB phase. 
The equation of state and the gas opacities are computed on-the-fly with the \texttt{\AE SOPUS} code \citep{MarigoAringer_09}, so as to  consistently follow  the variations in the chemical composition caused by mixing events and nucleosynthesis.  

Mass loss by stellar winds during the AGB phase is described with a two-regime scheme, first introduced in \cite{girardi10}:
\begin{enumerate}
    \item \label{item:i}  {\em Pre-dust mass loss (\mdotpre).} It applies as long as the conditions, mainly at lower luminosities and higher effective temperatures,  prevent the formation of dust grains and the development of a dust-driven wind. In this work we adopt the formalism presented by \citet{CranmerSaar_11}, which relies on the action of magnetic fields in the extended and cool chromospheres of red giants.  In our models the pre-dust mass loss typically takes place during the Early-AGB phase.
    \item \label{item:ii}  {\em Dust-driven mass loss (\mdotdust).} When AGB stars evolve to higher luminosities and large-amplitude pulsation develops, powerful stellar winds may be triggered through radiation pressure on dust grains which form in the extended and shocked atmospheres \citep{HoefnerOlofsson_18}.
    Here we adopt different mass-loss descriptions depending on the surface C/O ratio. As long as a star has C/O$< 1$, we use the \citet[][hereafter BL95]{bloecker95} formula, with an efficiency parameter $\etadust=0.03$. When the star attains  C/O$> 1$, as a consequence of the 3DU, we adopt the results of dynamical atmospheres models for carbon stars (hereafter CDYN) recently developed by \citet{mattsson10}, \citet{Eriksson_etal14}, and \citet{Bladh_etal_19}.
 \end{enumerate}
 
As detailed in \citet{marigo13}, the 3DU is modelled with a parametrized description that relies on three main characteristics:
\begin{enumerate}
    \item {\em Onset and quenching of the 3DU.} These are determined by a temperature criterion, $\Tbdred$, which is the minimum temperature that must be reached at the base of the convective envelope at the stage of post-flash luminosity maximum for a mixing event to occur.
    \item {\em Efficiency of the 3DU.} It is described by the standard parameter $\lambda=\Delta M_{\rm dred}/\Delta M_{\rm c}$, the fractional increment of the core mass during an inter-pulse period that is dredged up in the subsequent thermal pulse.
    In this work we adopt the new parametrization of $\lambda$ introduced by \citet{pastorelli19}. This scheme is designed to be: a) qualitatively consistent with full TP-AGB model calculations, and b) to contain suitable free parameters to perform a physically-sound calibration based on the observed C-rich star LFs. 
    The free parameters are: i) $\lambda_{\rm max}^{\ast}$, the maximum efficiency of the 3DU among all TP-AGB models; ii) $\widetilde{M}_{\rm c}$, the value of the core mass  for which $\lambda_{\rm max}^{\ast}$ is attained; and iii) $M_{{\rm c},\lambda=0}$, the value of the core mass above which the 3DU is no more active.  This latter  parameter is introduced to allow for the possibility that at larger core masses the average efficiency of the 3DU may decline, as indicated by some existing TP-AGB models \citep{VenturaDantona_09, Cristallo_etal_15}.
    \item {\em Chemical composition of the intershell.} Here we adopt the standard case described in \citet{marigo13}, where no overshoot is assumed at the convective boundaries, and the typical abundances of helium, carbon and oxygen are (in mass fraction):  $\mathrm{^4He/^{12}C/^{16}O \approx (0.70-0.75)/(0.25-0.20)/(0.005-0.01)}$.
\end{enumerate}

During the calibration cycle, each time a new set of TP-AGB tracks is computed for a new combination of parameters, the next step is the generation of a corresponding set of \parsec+\colibri\ stellar isochrones by means of the  \trilegal\ code, as detailed in \citet{marigo17}. 
We recall that \trilegal\ includes specific TP-AGB physical processes, such as the luminosity and temperature variations during the thermal pulse, the variations in the surface chemical compositions and spectral type, as well as the reprocessing of radiation by circumstellar dust. The photometry is calculated using extensive tables of bolometric corrections based on the spectral libraries by \citet{aringer09} for C-rich stars, and \citet{castelli03} and  \citet{aringer16} for O-rich stars. 

The synthetic photometry includes the effect of the circumstellar dust in mass-losing stars, as fully described by \citet{marigo08}.
Briefly, this approach is based on radiative transfer calculations across dusty envelopes \citep{groenewegen06, bressan98}, coupled with the scaling formalism first introduced by \cite{ElitzurIvezic_01} and a few key relations from the dust-growth model by \citet{FerrarottiGail_06}.
In \citet{pastorelli19} we made a few modifications to the dust treatment to improve the consistency of our simulations. We revised the abundance of some elements to follow the scaled-solar pattern of \citet{caffau11}, as in the evolutionary tracks, and we replaced the fitting relations to compute the condensation degree of carbon dust by \citet{FerrarottiGail_06}  with the results of dynamical atmosphere models by \citet{Eriksson_etal14}, which are also used to predict the mass-loss rates of C-rich stars during the dust-driven regime.  
For the present work we adopt tables of dust bolometric corrections based on spectra computed with the following dust mixtures: amorphous carbon (85 per cent) and SiC  (15  per  cent)  for  C-rich  stars,  and  silicates  for  O-rich stars  \citep{groenewegen06}. 

\section{Results}
\label{sec:results}

In this Section we present the results of our TP-AGB calibration for the LMC galaxy. 
We first describe our starting population synthesis model, and its performance compared to the observed LFs (see Sect.~\ref{ssec:start_model}). Then, we present the best-fitting model we find by acting on the 3DU parameters (Sect.~\ref{ssec:3DU}). 
Finally, in Section~\ref{ssec:new_opacity}, we discuss the effect of newly available line lists for modelling the spectra of C-rich stars, and their impact on the TP-AGB model calibration. 

\subsection{Starting model}
\label{ssec:start_model}

In \citet{pastorelli19} we identified two best-fitting models (the sets S\_07 and  S\_35) which reproduce the SMC infrared LFs in the 2MASS and \textit{Spitzer} filters, and the star counts for each class of AGB stars. Both models perform comparably well in recovering the observed LFs and CMDs, but with some preference towards the set S\_35 as it yields final masses for the white dwarfs (WDs) which are closer to the semi-empirical initial-final mass relation \citep[IFMR;][]{elbadry18, cummings18}.

\begin{figure}
\centering
\includegraphics[width=0.9\columnwidth]{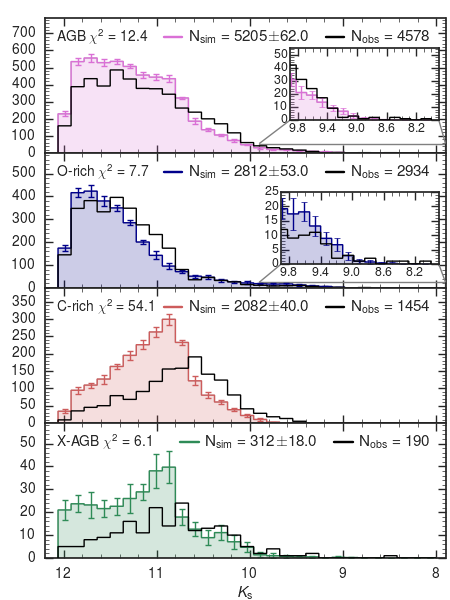}
\caption{Simulated mean \ks-band LFs based on S\_35 models (filled histograms in colour), compared to observed distributions (dark-line histograms), for the entire sample (top panel) and for the three main classes of TP-AGB stars (other panels). The error bars cover  1-$\sigma$ uncertainties resulting from 10 \trilegal\ realizations.  Each panel reports the  numbers of observed and synthetic stars, as well as the \chisq\ specific to each simulated LF.}
\label{fig:LFKsS35}
\end{figure}

We start by simulating the LMC photometry using \colibri\ TP-AGB evolutionary tracks with the S\_35 input prescriptions, summarized in Table~\ref{tab:mod_grid}. 
We show the performance of this set in Fig.~\ref{fig:LFKsS35}.
We note that while the number counts of the observed O-rich AGB stars are reasonably well reproduced, the faint end of the simulated LFs shows an excess which is compensated by a deficit at brighter magnitudes, i.e. \ks $\approx 11.5 - 10.8$~mag. Moreover, the most evident discrepancy is the overestimation of C-rich stars, especially at faint \ks\ magnitudes, by roughly 40 per cent. Similarly, a sizeable  excess is found in the simulated X-AGB stars.

\begin{figure}
\centering
\includegraphics[width=0.9\columnwidth]{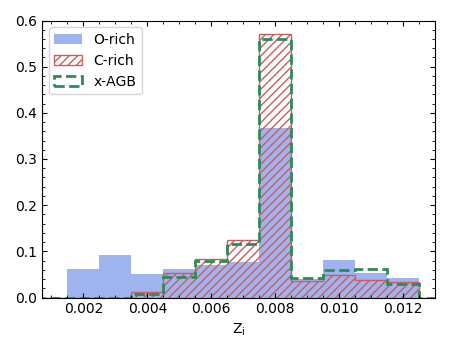}
\caption{Initial metallicity distributions of O-, C-, and X-AGB stars predicted by S\_35. The AMR is derived from the SFH recovery, and it is an input of our simulations.  }
\label{fig:LMC_Z_dist}
\end{figure}

In this respect, we recall that the set S\_35 was calibrated to reproduce the photometry of the SMC which, on average, has a lower metallicity than the LMC. In fact, the mean metallicity of SMC C-rich stars is $\Zini \approx 0.004$, whereas the bulk of LMC C-rich stars has $\Zini \approx 0.008$ (see Fig.~\ref{fig:LMC_Z_dist}). It follows that the excess of simulated C-rich and X-AGB stars in the LMC may be linked to the different metallicities that characterize the two galaxies. Hence, the need to check, and possibly revise, the starting TP-AGB calibration at higher metallicities.

In this work, we adopt the evolutionary tracks of the set S\_35, already calibrated in the SMC, for the metallicity range $\Zini = 0.0005 - 0.006$.
In addition, for higher metallicities, $\Zini = 0.008, 0.01, 0.014, 0.017$, and $0.02$, we compute around 70 tracks with initial masses in the range between 0.5 and 5-6 \Msun, for each new combination of input parameters.
This choice is also motivated by the fact that the predicted initial metallicity distributions of all classes of AGB stars show a pronounced peak at $\Zini \approx 0.008$, as shown in Fig.~\ref{fig:LMC_Z_dist}.
We emphasize that the metallicity distributions are based on the AMR from the SFH recovery. As such, the present calibration only probes metallicities as high as $\Zini \approx 0.012$. 
We note that in our \trilegal\ simulations we use a set of \colibri\ tracks covering the entire metallicity range expected for the LMC population. 

\begin{table}
\centering 
\caption{Third dredge-up parameters of the TP-AGB sets presented in this work. \label{tab:mod_grid}}
\begin{tabular}{cccc}
\hline
\multirow{2}{*}{SET}   & \multicolumn{3}{c}{Third dredge-up}  \\
         & $\lambda_{\rm max}^{\ast}$ &
      $\mathrm{\widetilde{M}_c}[\Msun]$      &   
      $\mathrm{M_{c,\lambda=0}[\Msun]}$ \\ 
      \hline
S\_35   & 0.7  & 0.60 & 1.00 \\ 
S\_36   & 0.7  & 0.70 & 1.00 \\ 
S\_37   & 0.5  & 0.70 & 1.00 \\ 
\hline
\end{tabular}
\end{table} 

\subsection{Characterizing the 3DU in the LMC}
\label{ssec:3DU}

\begin{figure*}
  \includegraphics[width=0.9\textwidth]{{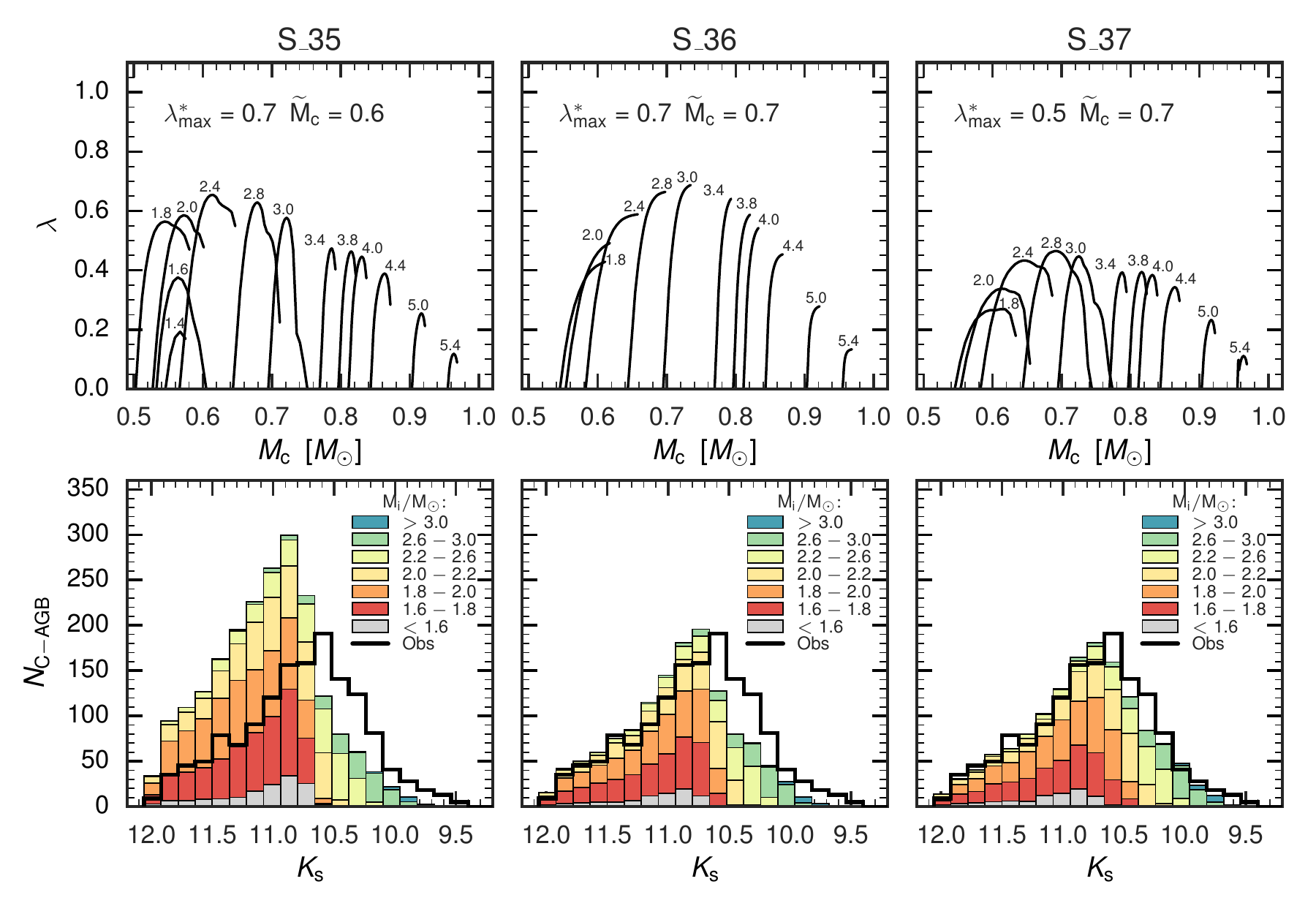}}
    \caption{Top rows of each panel: efficiency of the 3DU ($\lambda$) as a function of the core mass $M_\mathrm{c}$ of a few selected evolutionary tracks with $\Zini=0.008$ and initial mass as labelled in the figure. Bottom rows of each panel: observed (black histograms) and simulated C-rich LFs as derived from the corresponding above sets of models. The synthetic LFs are shown as stacked histograms to highlight the contribution of each initial mass bin to the LF as indicated in the legend.}
    \label{fig:lambda_plot_1fig}   
\end{figure*}


\begin{figure*}
\begin{center}
\begin{minipage}[t]{.49\textwidth}
\centering
\includegraphics[width=0.9\columnwidth]{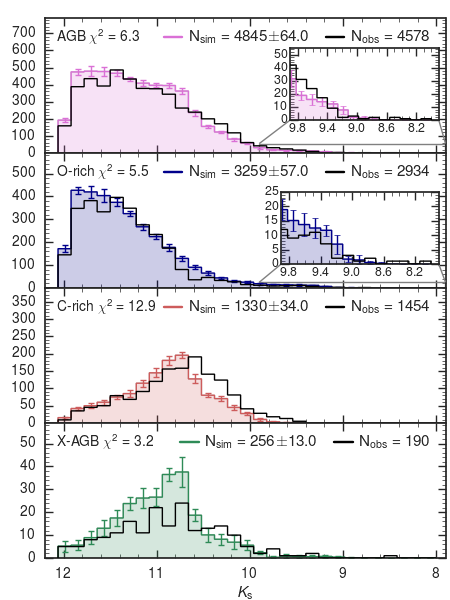}
\caption{Same as Fig.~\ref{fig:LFKsS35}, but for S\_36 in which we increase the 3DU parameter $\mathrm{\widetilde{M}_c}$ from 0.6 to 0.7~\Msun. }
\label{fig:LFKsS_LMC_36}
\end{minipage}%
\hspace{2mm}
\begin{minipage}[t]{.49\textwidth}
\centering
\includegraphics[width=0.9\columnwidth]{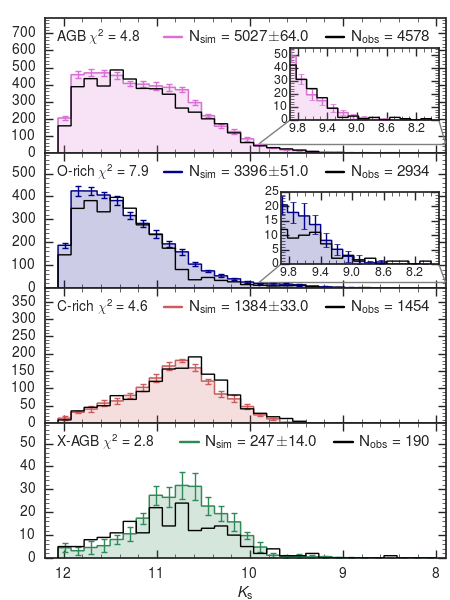}
\caption{Same as Fig.~\ref{fig:LFKsS_LMC_36}, but for S\_37 in which $\lambda_{\rm max}^{\ast}$ is lowered from 0.7 to 0.5. }
\label{fig:LFKsS_LMC_37}
\end{minipage}%
\end{center}
\end{figure*}

As a first attempt to reduce the number of low-mass faint C-rich stars, we compute the \colibri\ set S\_36 in which we keep the same mass-loss prescription as in S\_35, whereas we increase the 3DU parameter $\mathrm{\widetilde{M}_c}$ from 0.6 to 0.7~\Msun. To better appreciate the global effect, in Fig.~\ref{fig:lambda_plot_1fig} we show the efficiency of the 3DU as a function of the core mass ${{M}_c}$ for a few TP-AGB evolutionary models with \Mini $\geq 1.4~\Msun$ (top panels), and the corresponding simulated C-rich star LFs (bottom panels).  In each panel $\mathrm{\widetilde{M}_c}$ corresponds to the value of the core mass for which $\lambda$ attains the maximum value, $\lambda_{\rm max}^{\ast}$.
The main effect of increasing this parameter is to delay the onset of the 3DU at larger core masses (in addition to the temperature criterion) in all TP-AGB models; in particular the occurrence of the mixing events is even prevented in stars with $\Mi < 1.65\, \Msun$ at $\Zini = 0.008$. This depopulates the faint wing of the C-rich LF, leading to a  better agreement of the models with the observed data.

At the same time, increasing $\mathrm{\widetilde{M}_c}$ shifts the maximum 3DU efficiency to stars of larger mass, from $\Mini \approx 2.4~\Msun$ in set S\_35 to $\mini \approx 3.0~\Msun$ in set S\_36.  
The reduction in the number of C-rich stars is significant, and the simulated C-rich LF agrees better with the observed one.  

The set S\_36 improves in the simulated O-rich LF as well (see Fig.~\ref{fig:LFKsS_LMC_36}). In particular, it reduces the deficit of O-rich stars in the bright wing of the LF ($\ks \lesssim 11.4$~mag). 
This is the consequence of delaying the onset of the 3DU in intermediate mass-stars ($\Mi \approx 2-3~\Msun$), so that the O-rich stages extend over brighter luminosity bins. 

Though the improvement obtained with the set S\_36 is already appreciable, the calibration can be further refined  to fill the deficit of C-rich stars in the bright-end of the simulated LF, by stretching the distribution to slightly brighter magnitudes, and to reduce the excess of C-rich and X-AGB stars fainter than $\ks \approx 10.7$~mag. Following the analysis carried out by  \citet{pastorelli19}, the natural step is to decrease the $\lambda_{\rm max}^{\ast}$ parameter. We compute the set S\_37, in which $\lambda_{\rm max}^{\ast}$ is lowered from 0.7 to 0.5.  The net effect is a moderate reduction of $\lambda$ in all stellar models (see top right panel of Fig.~\ref{fig:lambda_plot_1fig}), with the consequence that the transition to C/O $> 1$, and the subsequent carbon-rich phases, take place at somewhat brighter magnitudes, resulting in a small shift of both C-rich and X-AGB distributions (bottom  right panels of Figs.~\ref{fig:lambda_plot_1fig} and \ref{fig:LFKsS_LMC_37}). The improvement obtained moving from S\_36 to S\_37 is quantitatively measured by comparing the \chisqlf\ of the simulated distributions, which decreases from 12.9 to 4.6 for the C-rich stars.

With the set S\_37 we reach a satisfactory agreement with the observations. We obtain the lowest \chisqlf\ values for the entire AGB sample, and for the classes of O-, C- and X-AGB, at the same time. 
The total AGB number counts are matched within 10 per cent, while the predicted C-rich number counts are within 5 per cent. The residual excess of simulated O-rich stars, $\approx 15$ percent, can be reduced to $\approx 10$ percent by taking into account the mismatch in the predicted $\jks$ colours, as discussed in Sect.~\ref{ssec:data_model_comp}. 
Despite the significant improvement in both the number counts and the shape of the X-AGB LF with respect to the starting model, a residual excess of simulated stars is present at brighter magnitudes, i.e. $\ks \lesssim 11$~mag. 
In this respect, we emphasize that our calibration is focused on reproducing the bulk of O- and C-rich AGBs, and it is not affected by the low numbers of X-AGB sources. Furthermore the X-AGB stars are photometrically selected, and their predicted magnitudes are largely dependent on the adopted circumstellar dust prescriptions. 

In summary, our analysis leads us to  conclude that the 3DU in TP-AGB stars of the LMC should be somewhat less efficient than in TP-AGB stars of the SMC. This result is consistent with the qualitative trends predicted by full TP-AGB models \citep[e.g. ][]{karakas02, VenturaDantona_09, Cristallo_etal_11}, and earlier calibration studies \citep[e.g.][]{marigo07}.
 
\subsection{Effects of new opacity data in carbon star spectra}
\label{ssec:new_opacity}

\begin{figure}
    \centering
    \includegraphics[width=0.9\columnwidth]{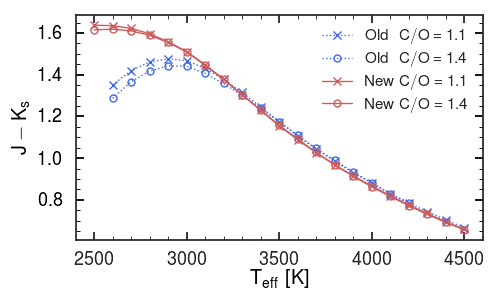}
    \caption{Predicted \jks\ colour as a function of effective temperature calculated from \textsc{comarcs} models with a surface gravity of $\rm log(g~[cm/s^2]) = 0.0$, a stellar mass of $M=1 \Msun$ and solar abundances except for carbon. Two C/O ratios (1.1 - crosses, 1.4 - circles) are shown. The results are based on the standard grid by Aringer et al. (2016, old, blue dotted lines) and on computations with the new opacities for C$_2$H$_2$ and C$_2$ (new, red solid lines).} 
    \label{fig:newCdata_JK}
\end{figure}
\begin{figure*}
    \centering
    \includegraphics[width=0.9\textwidth]{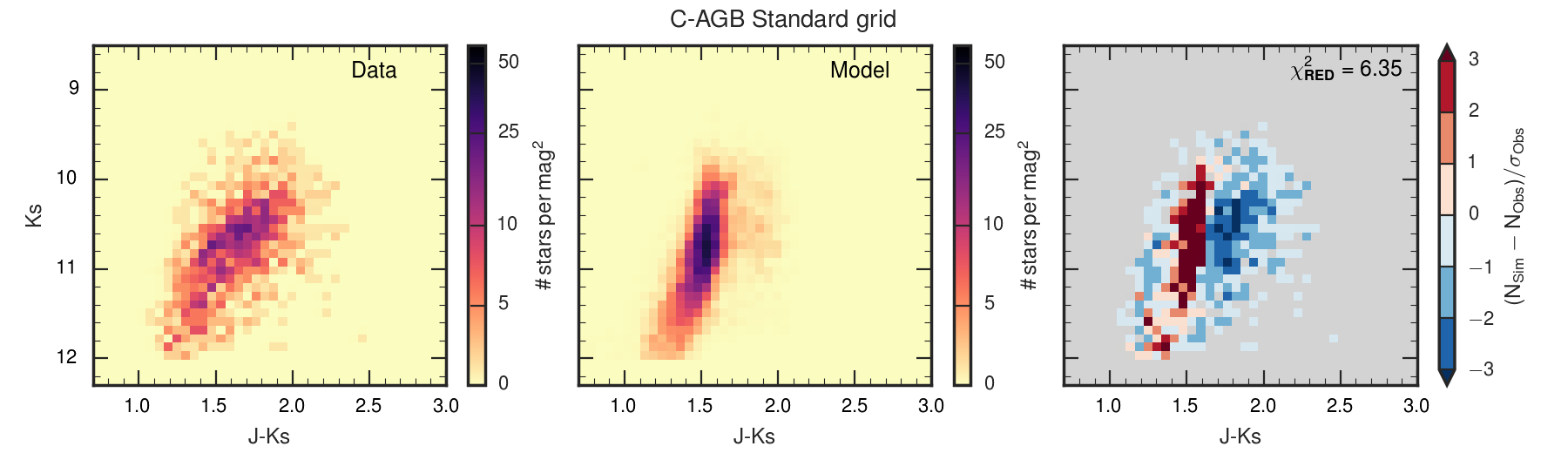}
    \includegraphics[width=0.9\textwidth]{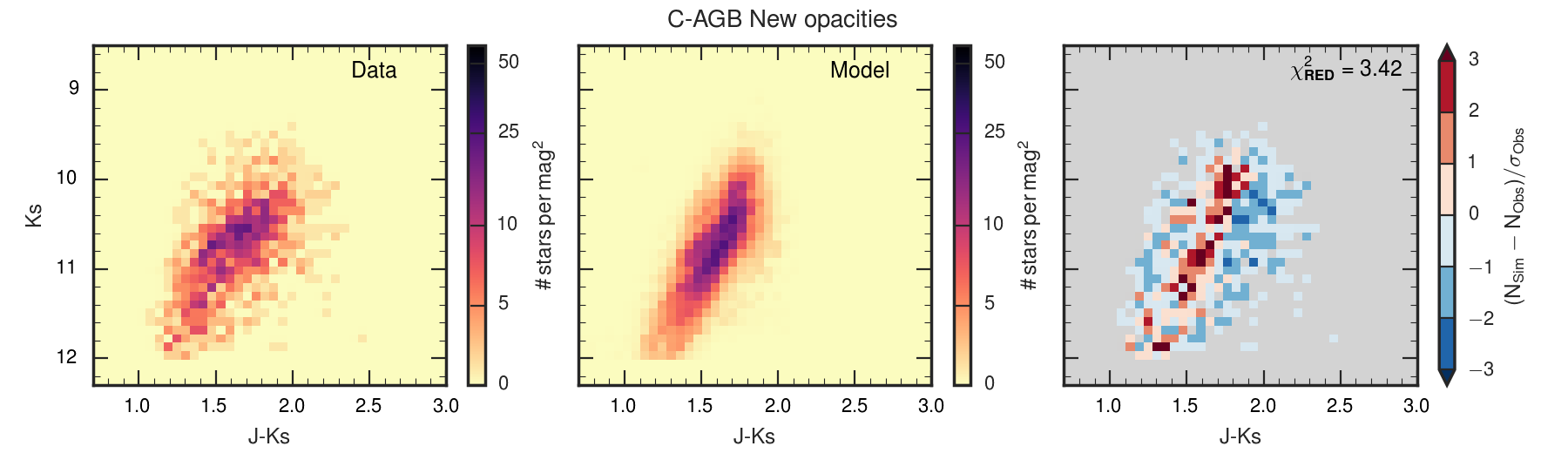}    
    \caption{Comparison between the C-rich Hess diagrams calculated on the basis of the standard spectral grid by Aringer et al. (2016, top panels) and the new grid based on new molecular opacity data (bottom panels). The results from the  best-fitting model S\_37 (middle panels) are shown together with the observed diagram (left panels). The map of the fractional difference between observed and simulated number counts, normalized to 1$\sigma$ of the observed counts, is also plotted (right panels), along with the values of the \chisq.}
    \label{fig:new_opacity_hess}
\end{figure*}

In recent years, new molecular linelists have  become available for some of the species important to model carbon star spectra. Compared to the original grid of \citet{aringer09} the inclusion of these data in the calculation of atmospheric models and observable properties causes significant changes in the pressure-temperature structures, synthetic spectra and photometric colours. The largest effects are due to new linelists for C$_2$ \citep{Yurchenko2018} and C$_2$H$_2$ \citep{Chubb2020}, computed as part of the \mbox{ExoMol} project \citep{ExoMol_database}. Updated opacities were also published for HCN \citep{Barber2014}, CH \citep{Masseron2014} and the lower levels of CN \citep{Brooke2014,Sneden2014}. However, the latter will only give rise to small changes in the overall energy distribution. The most important differences concerning the \jks\ indices of carbon stars are caused by the new C$_2$H$_2$ data, which produce a much lower opacity of this species in the range between 1 and 2~$\mu$m resulting in considerably redder colours.

Fig.~\ref{fig:newCdata_JK} compares the predicted \jks\ colours as a function of effective temperature between the previous sets of \textsc{comarcs} models from \citet[][and including updates from  \citealt{aringer16}]{aringer09} and the new set of
models that include the new opacity data for C-rich stars. In this comparison, models are limited to a single value of surface gravity ($\log g=0$), and refer to dust-free stars. The \jks\ colours obtained with the two opacity sets are the same for relatively high effective temperatures. However, for \Teff\ values below 3200~K, the \jks\ colours predicted by the new models become about 0.2~mag redder than those from \citet{aringer09}. 

The net effect of the new opacities on populations of AGB stars will depend on the distribution of \Teff, \logg, carbon excess, and dust properties of the model AGB stars. Unfortunately, the grid of new \textsc{comarcs} models is not yet complete enough to be implemented straight away in \trilegal, or to model all these dependencies in a consistent way. 
However, the computed grid is large enough to assess the differences in the predicted magnitudes between the two versions of the same simulation. Therefore, we derive a relation that yields the corrections to be applied to the magnitudes given by the standard set of spectra in order to recover the results obtained with the new opacity data. This is based on \textsc{comarcs} models at solar metallicity, but the effect for lower metallicity C-rich stars is expected to go in the same direction. The correction contains a dependence on both effective temperature and carbon excess. We apply it to each C-rich star of the S\_37 simulation.
The main effect of the new opacity data is to predict brighter \ks\ magnitude for stars with effective temperatures below 3200 K. 
Adopting this correction leads to a better fit to the observed bright wing, as proved by the lower value of the \chisq\ which decreases  from 4.6 down to 2.1 for the set S\_37.
The most evident improvement is found in the predicted near-IR colours. In Fig.~\ref{fig:new_opacity_hess} we show the observed and simulated \cmd{\ks}{J}{\ks} Hess diagrams for the C-rich population. For the \citet{aringer09} models, the \jks\ colours reach a maximum value of $\approx$ 1.6 mag, with the bulk of C-rich stars aligning along an almost vertical structure at brighter magnitudes. Conversely, the bulk of the observed C-rich population shows redder \jks\ colours at brighter \ks\ magnitude. The observed behaviour is now better reproduced if we adopt the correction derived from the new opacity data. The slope of the simulated C-rich sequence shows a bending towards redder \jks\ colours similar to the observations. The improvement in the \jks\ vs. \ks\ diagram can also be appreciated by considering, for each cell of the Hess diagram, the difference between the number of simulated and observed number counts, relative to 1 $\sigma$  of the predicted distribution (Fig.~\ref{fig:new_opacity_hess}).
The value of the \chisq\ is reduced from 6.35 down to 3.42 as we move from the old to the new simulations. 

To test the impact of the new opacity on our previous TP-AGB calibration \citep{pastorelli19} we perform the same kind of test on the SMC and apply the correction to the simulations calculated with the best-fitting set S\_35. We find no significant changes in the \ks-band LFs, nor in the \cmd{\ks}{J}{\ks} CMDs. The reason is that C-rich stars in the SMC have, on average,  higher effective temperatures compared to the LMC, and this characteristic makes the correction smaller for the SMC, as shown in Fig.~\ref{fig:newCdata_JK}.

\subsection{Observed and best-fitting \cmd{\ks}{J}{\ks} CMDs}
\label{ssec:cmd_obs_sim}

\begin{figure*}
    \centering
    \includegraphics[width=\textwidth]{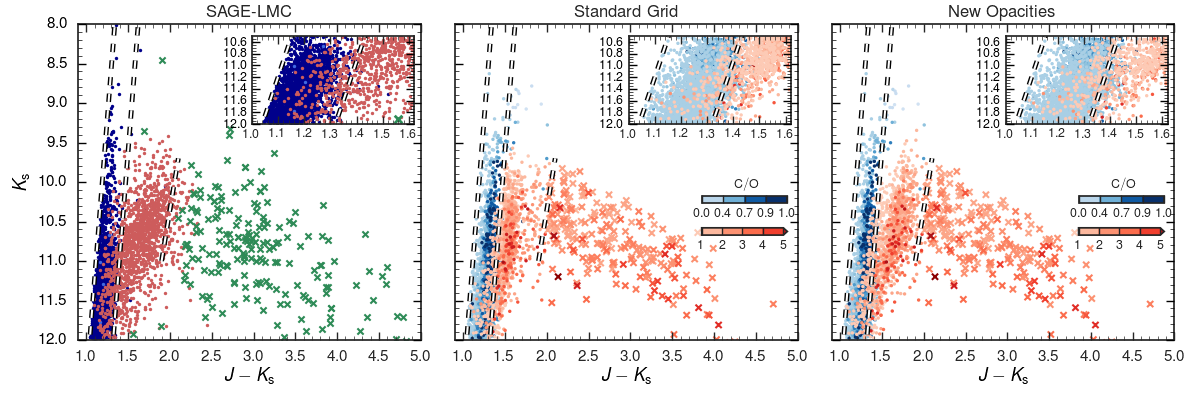}
    \caption{Left panel: observed \cmd{\ks}{J}{\ks} CMD with stars colour-coded according to the their classification in O-rich (blue), a-AGB (light blue), C-rich (red) and X-AGB (green crosses). Middle and right panels: simulated CMDs from the best-fitting set S\_37 calculated with the standard spectral grid by \citet{aringer16} and the grid based on new molecular opacity data. The simulated stars are colour-coded according to the predicted \co\ ratio. The insets show the CMD region where the O-rich and C-rich stars cannot be distinguished in the \cmd{\ks}{J}{\ks} CMD.
    The dashed lines are the photometric criteria used to select the observed O- and C-rich stars \citep{boyer11}. A third line shows the approximate separation between C- and X-AGB stars.}
    \label{fig:cmd_obs_S37}
\end{figure*}

We show the comparison between the observed and the simulated \cmd{\ks}{J}{\ks} CMDs from the best-fitting model S\_37 in Fig.~\ref{fig:cmd_obs_S37}. In the observed CMD, the stars are colour-coded according to their classification in C-, O-, X-, and a-AGB stars, whereas the synthetic stars are colour-coded according to the predicted C/O ratio. As described in Sect.~\ref{ssec:new_opacity}, the colours of C-rich stars are in better agreement with the observations, when the effects of new opacity data are taken into account. In this case, the distribution of simulated C-rich stars matches the observed \cmd{\ks}{J}{\ks} colours, in particular for stars brighter than $\ks \approx 11$~mag, which extend to the line that approximately separates C- and X-AGB stars. The insets highlight the CMD region in which C- and O-rich stars cannot be distinguished with the 2MASS photometry alone. The predicted distribution of C- and O-rich stars in this region is in fair agreement with the observed one. A slight shift towards redder colours is visible in the distribution of faint O-rich stars. As discussed in Sect.~\ref{ssec:data_model_comp}, this discrepancy does not affect the results of our calibration.

\subsection{2MASS and {\em Spitzer} luminosity functions}
\label{ssec:lfs_tmass_spitzer}

\begin{figure*}
\centering
\includegraphics[width=\textwidth]{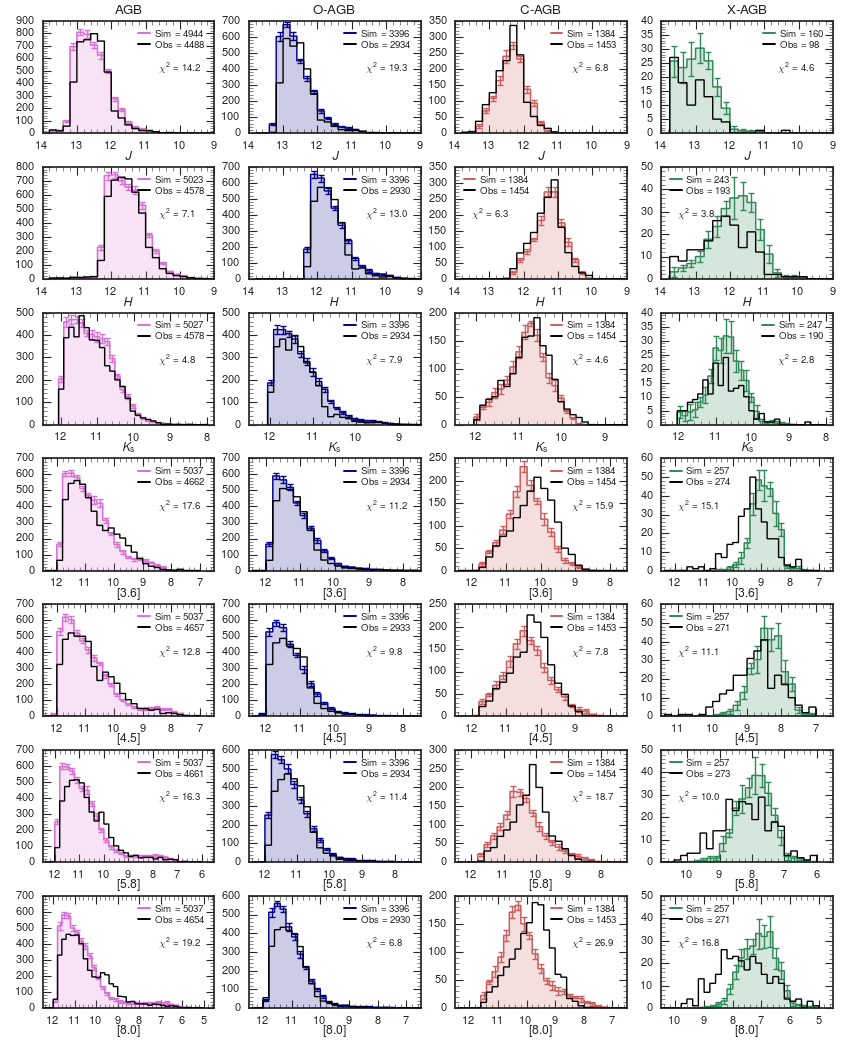}
\caption{Comparison between the synthetic LFs obtained from the best-fitting model S\_37 and the observed LFs in the 2MASS and Spitzer filters, going from shorter (top panels) to longer wavelengths (bottom panels). }
\label{fig:sage_lfs_S37}
\end{figure*}

We test the performance of the best-fitting set S\_37 by comparing the predicted and observed LFs in the 2MASS and \textit{Spitzer} bands available in the SAGE-LMC catalogue.
Fig.~\ref{fig:sage_lfs_S37} shows such comparison for the AGB sample and for each class of AGB stars in the 2MASS $J$, $H$, \ks\ filters, and the \textit{Spitzer} [3.6], [4.5], [5.8], [8.0] filters.
In general, the synthetic LFs are in agreement with the observed ones, with some exceptions. Specifically, the predicted C- and X-AGB LFs in the \textit{Spitzer} bands are shifted to fainter and brighter magnitudes respectively. 
As discussed by \cite{pastorelli19} for the SMC, where we the same kind of discrepancy is found (see their figure 21), the differences in the X-AGB LFs do not impact our results as the percentage of X-AGB stars is less than 6 per cent of the total AGB sample. However, these discrepancies may be used to improve the circumstellar dust treatment and to test the mass-loss prescriptions in the advanced stages of TP-AGB evolution. We plan to address this point in a future study. 

\section{Discussion}
\label{sec:discussion}

In the following we discuss the main relevant implications expected from the present calibration. 
We discuss the predictions of the set S\_37 in terms of initial masses of C-rich stars, mass-loss rate distributions, TP-AGB lifetimes, and contribution of TP-AGB stars to the integrated luminosity. 

\subsection{The domain of carbon stars in the LMC}
\label{ssec:cstars}
\begin{figure*}
\centering
\includegraphics[width=0.8\textwidth]{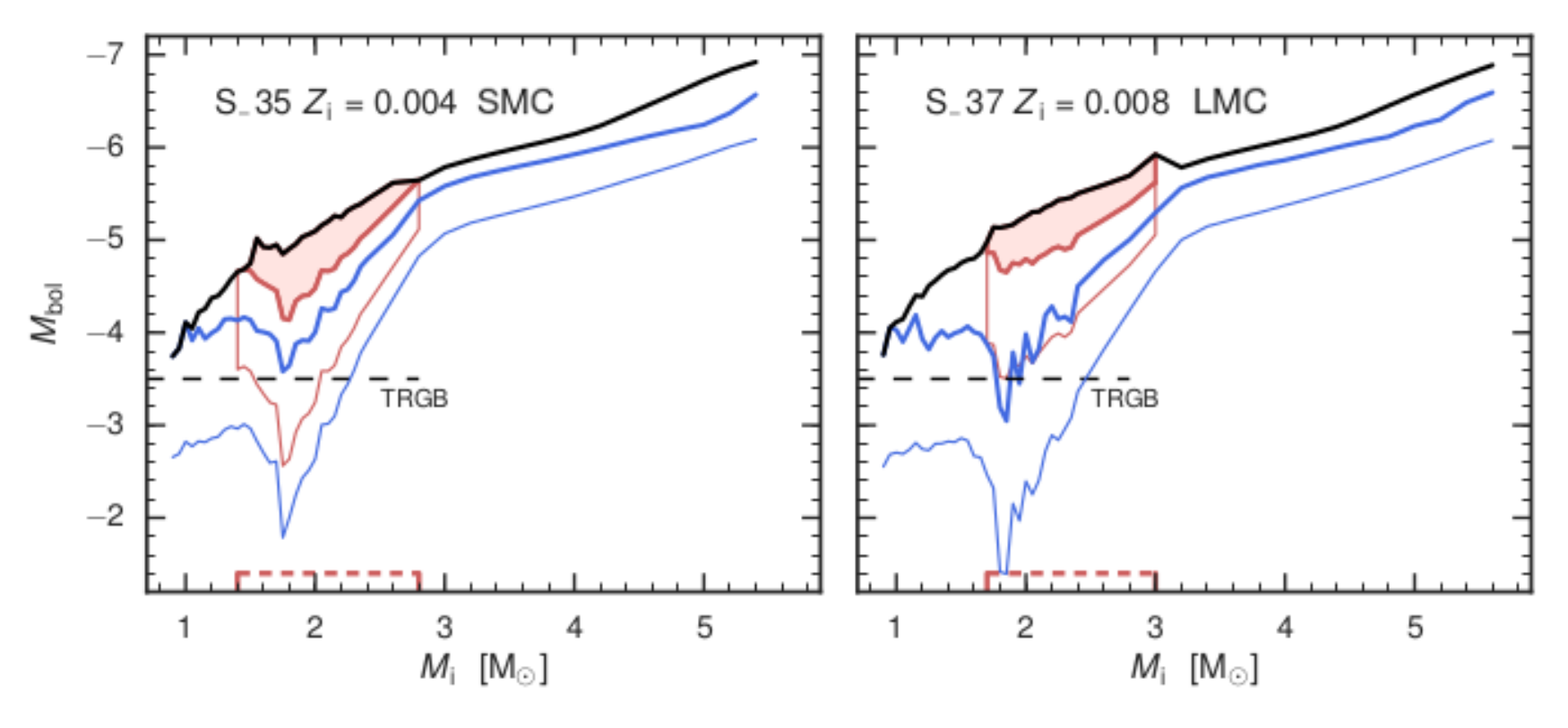}
\caption{Bolometric magnitudes as a function of \Mini\ for a few relevant transition stages: the first TP (blue), the transition from the O-rich to the C-rich domain (red), and the AGB tip (black). 
The thick solid lines connect the quiescent stages just prior to the occurrence of TPs, while thin solid lines correspond to the faintest luminosities reached during the post-flash low-luminosity dips. 
Results are shown for the TP-AGB sets S\_35 at $\Zini = 0.004$, calibrated in the SMC, and S\_37 at $\Zini=0.008$ (LMC calibration). The range of initial masses for the formation of C-rich stars is shown by the horizontal red-dashed lines.}
\label{fig:mbol_tr_lum}
\end{figure*}

Fig.~\ref{fig:mbol_tr_lum} shows the predicted ranges of initial masses and bolometric magnitudes of C-rich stars for our best-fitting models, namely the set S\_35 with $\Zini = 0.004$ based on the SMC calibration and the set S\_37 with $\Zini = 0.008$ based on the LMC calibration.
The C-rich star domain extends from $\Mini \approx 1.4~\Msun$ to $\Mini \approx 2.8~\Msun$ at $\Zini = 0.004$ and from $\Mini \approx 1.7~\Msun$ to $\Mini \approx 3~\Msun$ at $\Zini = 0.008$.
The minimum mass for producing carbon stars is found to decrease with decreasing metallicity, a finding that supports theoretical trends in the literature \citep[e.g.][]{marigo07, Cristallo_etal_11, Cristallo_etal_15, ventura13}. 

\begin{figure}
\centering
\includegraphics[width=0.47\textwidth]{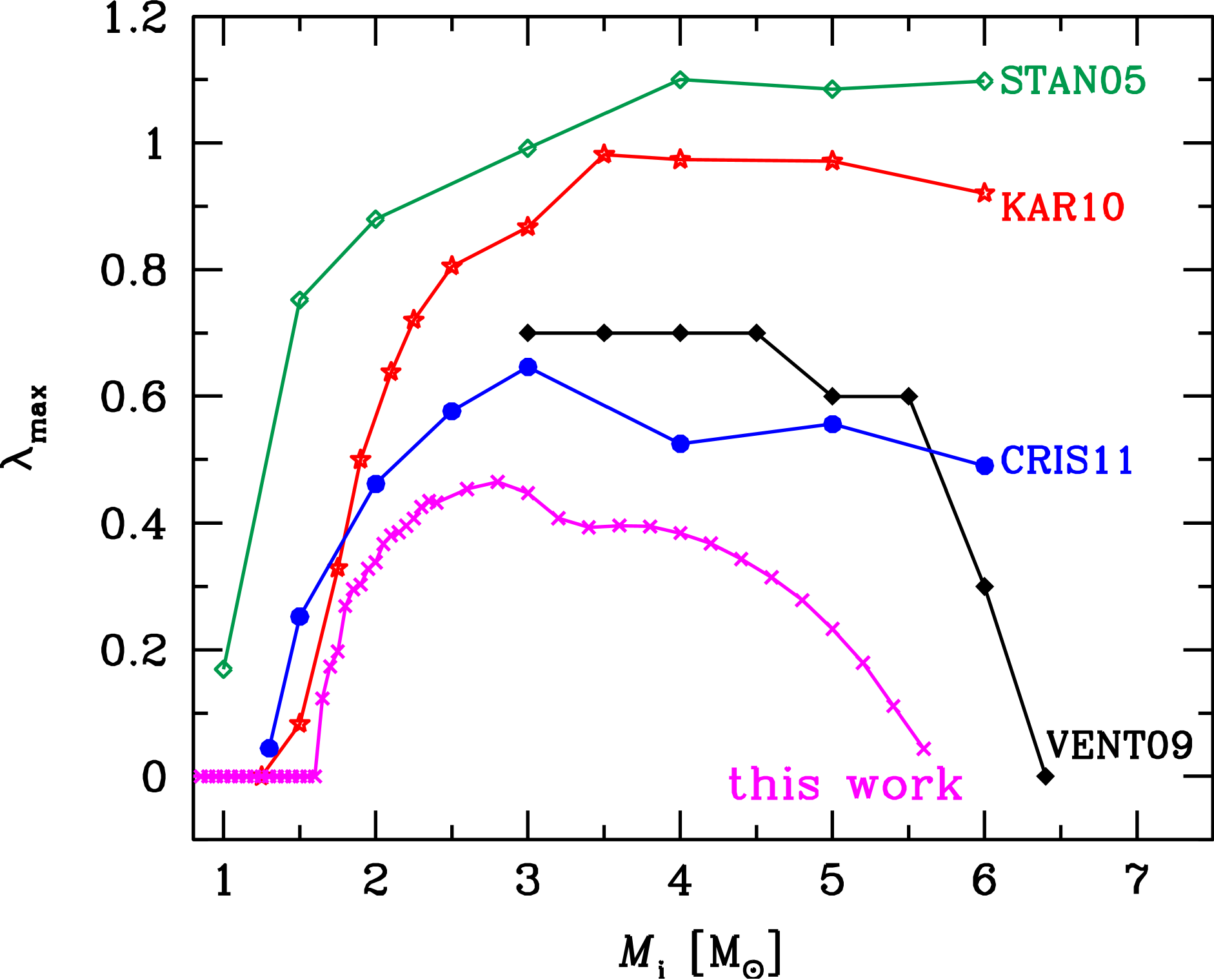}
\caption{Maximum efficiency of the 3DU as a function of the initial stellar mass, predicted  by AGB models available in the literature, namely: \citet{Stancliffe_etal_05} (green), \citet{Karakas10} (red), \citet{Cristallo_etal_11} (blue), \citet{VenturaDantona_09} (black). Our calibrated values  are shown for comparison (magenta).
All cases refer to $\Zini=0.008$, except for \citet{VenturaDantona_09} models that correspond to $\Zini=0.006$.
Note the substantial differences from author to author for $\Mi > 2~\Msun$.}
\label{fig:lambvar}
\end{figure}

It is now useful to compare the 3DU properties predicted by available  full AGB models in the literature with the results of our calibration, focusing on the metallicity $\Zini=0.008$ (or similar) which characterizes most of the  carbon stars in the LMC (see Fig.~\ref{fig:LMC_Z_dist}).
As shown in Fig.~\ref{fig:lambvar}, the efficiency of the 3DU is still affected by significant differences from author to author, mostly evident for $\Mi \gtrsim 2~\Msun$. Models by \citet{Stancliffe_etal_05} and \citet{Karakas10} correspond to the largest $\mathrm{\lambda_{max}}$  which approaches $\approx 1$ for $\Mini \gtrsim 3~\Msun$, while considerably lower values ($\mathrm{\lambda_{max}} \simeq 0.5-0.7$) are predicted by \citet{Cristallo_etal_11} and  \citet{VenturaDantona_09}. In particular, at increasing $\Mini$, the parameter  $\mathrm{\lambda_{max}}$  declines down to zero in \citet{VenturaDantona_09} models, contrarily to the findings of \citet{Stancliffe_etal_05} and \citet{Karakas10}.
Our calibrated relation for $\mathrm{\lambda_{max}}$ presents a trend closer to the results of  \citet{Cristallo_etal_11} and \citet{VenturaDantona_09} but shifted to lower values.
 
Concerning the initial mass range of carbon stars the situation is illustrated in Table~\ref{tab:crange}.
Let us denote with $\mathrm{M_{Cstar}^{min}}$ and $\mathrm{M_{Cstar}^{max}}$
the minimum  and maximum initial mass for carbon star formation.
We see that  $\mathrm{M_{Cstar}^{min}}$ at $\Zini =0.008$ varies from $1~\Msun $ to $2.4 ~\Msun$. This is a notable scatter since  the difference in mass translates into a wide age range, from $\approx 9.4$~Gyr to $\approx 0.7$~Gyr.
The upper limit $\mathrm{M_{Cstar}^{max}}$  is found to vary between $\approx 3~\Msun$ to $\approx 5~\Msun$, which corresponds to  an age interval from $\approx 0.4$~Gyr to $\approx 0.1$~Gyr.  We note that our results for  $\mathrm{M_{Cstar}^{max}} \simeq 3 \Msun$  agree with the predictions of  \citet{Stancliffe_etal_05}, \citet{Cristallo_etal_11}  and \citet{dellagli15}. 
 
In summary, our calibration indicates that the 3DU in TP-AGB LMC stars (with $\Zini=0.008$) has an efficiency not exceeding $\lambda \approx 0.5$ at all initial masses. This result conflicts with the predictions of some AGB models \citep[e.g.,][]{Stancliffe_etal_05, Karakas10} in which $\lambda$ is much higher.
The bulk of C-stars in the LMC should have ages between $\approx 1.7$ ($\mathrm{M_{Cstar}^{min}} \simeq 1.7~\Msun$) and  $\approx 0.4$~Gyr ($\mathrm{M_{Cstar}^{min}} \simeq 3.0~\Msun$).
 
Finally we note that C-stars with  $\Mi < 1.7~\Msun$ are expected to be present within the AGB population of the LMC, but these should have a lower metallicity ($\Zini < 0.008$) according to the AMR. Our best-fit simulations indicate  that C-stars exist down to $\Mi \approx 1.4~\Msun$ and $\Zini \approx 0.004$.
Similarly, the most massive carbon stars are found to have $\Mi \approx 3.2~\Msun$  and metallicity $\Zini \approx 0.01$.

 \begin{table}
     \centering
     \caption{Initial mass limits for C-star formation}
     \begin{threeparttable}
     \begin{tabular}{lccc}
     \hline
     \noalign{\smallskip}
     Reference & $\mathrm{M_{Cstar}^{min}}$ & $\mathrm{M_{Cstar}^{max}}$ & \Zini\\
         \noalign{\smallskip}
      & $\mathrm{[M_{\odot}]}$ & $\mathrm{[M_{\odot}]}$ \\
     \hline
     \noalign{\smallskip}
     \citet{Stancliffe_etal_05} & 1.00 & 3 & 0.008\\
     \citet{WeissFerguson_09} & 1.00 & 5 & 0.008\\
     \citet{Karakas10} & 1.75 & 4 & 0.008 \\
     \citet{Cristallo_etal_11} & 1.50 & 3 & 0.008 \\
     \citet{dellagli15} & 1.25 & 3 & 0.008\\
     \citet{Pignatari_etal_16} & 1.65 & 4 & 0.01\\
     \citet{Choi_etal_16} & 2.4 & 3.2 & 0.008\tnote{*}\\
     Our calibration & 1.70 & 3 & 0.008 \\
     \noalign{\smallskip}
     \hline     
     \end{tabular}
     \begin{tablenotes}\footnotesize
     \item[*] The MIST models for a metallicity [Fe/H]$=-0.25$ are obtained through the web-interface at http://waps.cfa.harvard.edu/MIST/interp\_tracks.html
     \end{tablenotes}
     \end{threeparttable}
     \label{tab:crange}
 \end{table}
 
At this point, one may wonder if these mass ranges agree with those inferred from the presence of C-rich stars in LMC star clusters. For instance, the classical paper by \citet{frogel90} mentions 3--5 \Msun\ as the maximum mass for the formation of C stars in the LMC, and does not find any C-rich stars among the oldest clusters corresponding to type VII in the \citet{swb} classification (ages larger than a few Gyr). These observations suggest constraints to the minimum and maximum masses for the formation of C-rich stars; however they are based on rough and largely outdated estimates of ages and turn-off masses, \Mto, for the hosting star clusters. As already mentioned in \citetalias{pastorelli19}, this analysis is better done using cluster ages based only on isochrone fitting applied to high-quality photometry from the Hubble Space Telescope (HST). A quick revision of the tables in \citet{girardi07} -- which includes the additional list of cool giants in MC clusters from \citet{vanloon05} -- reveals that:

\begin{enumerate}
    \item NGC~1850, with $\Mto>5.5$~\Msun\ (and hence evolved stars in which HBB is fully operating to prevent the formation of C-rich stars), is the youngest LMC cluster to contain a C-rich star within its core radius. It is followed by the 200-Myr old LMC cluster HS~327-E ($\Mto\sim4~ \Msun$) which contains both an obscured carbon star and an  OH/IR star \citep{vanloon01}.
    \item The next in the list of youngest clusters to contain C stars is either NGC~1987 or NGC~2209, with \Mto\ slightly below 2 \Msun, and both with several C-rich stars;
    \item the oldest one with a significant population of C-rich stars (about 10) is NGC~1978 with 2.2 Gyr and $\Mto\approx1.45$~\Msun\ \citep[][and Chen et al., in preparation]{mucciarelli07};
    \item The oldest one with a {\em modest} population of C stars (just one) appears to be  NGC~2121, for which HST data provides ages between 2.5~Gyr \citep{martocchia19} and 3.2~Gyr \citep{rich01, li19}, and hence a turn-off mass just slightly smaller than NGC~1978; 
    \item The age sequence is followed by the well-known gap in the history of cluster formation in the LMC -- that means, a lack of clusters with ages from this limit up to 9 Gyr \citep[see e.g.][for a more complete discussion]{sarajedini98, rich01, piatti02, piatti02b} -- and then by a handful of very old and populous globular clusters which do not contain (nor are expected to contain) C-rich stars.
\end{enumerate}
 
The C star in NGC~1850 deserves some discussion, because of its unusual age, and because this cluster is representative of more recent changes in the interpretative scenario of the CMDs of Magellanic Cloud star clusters. First of all, one might wonder if its C star \citep[2MASS J05085008-6845188, cf.][]{boyer13} is a chance alignment of a field star, or a true cluster member which already ended its HBB phase, and then underwent a late transition to the C-rich phase after losing a significant fraction of its initial stellar envelope. This latter scenario applies to the obscured carbon star in HS~327-E, which also contains the OH/IR star IRAS 05298-6957 \citet{vanloon01}. The main problem in this interpretation, however, is that the C star in NGC~1850 has a spectrum typical of most optically-visible C stars in the LMC field, with a 2MASS $\ks=10.818$  and $\jks=1.651$. Second, it is worth noticing that there are indications of a significant dispersion in the rotational velocities of stars at the turn-off of NGC~1850 \citep{bastian17}. This causes significant uncertainty in the initial masses inferred for its post-main sequence stars. For instance, \citet{bastian17} estimate $\log(\mathrm{age/yr})=7.9$ and a turn-off mass of $4.86$~\Msun, while \citet{johnston19} suggest that $\log(\mathrm{age/yr})$ can be as high as 8.27 and turn-off masses be as low as $1.9-4.0$~\Msun. 

If we disregard the single C-rich star in NGC~1850 and the obscured one in HS~327-E, we are left with firm detections of C-rich stars only in LMC clusters with turn-off masses in the narrow interval from about $1.4$ to 2~\Msun. The observed interval is probably very much affected by the natural scarcity of TP-AGB stars in the youngest clusters with higher turn-off masses, and might be affected by the scarcity of star clusters in the age range corresponding to the lowest turn-off mass. 
Star clusters in the LMC, such as NGC~1978 and NGC~2121, indicate that the minimum mass to form C-rich stars should be around $\Mi \approx 1.45~\Msun$, which is lower than the predicted limit of $\Mi \approx~1.7~\Msun$ at $\Zini=0.008$. However, the discrepancy is only apparent since the metallicity of NGC~1978  is $\Zini \approx 0.005$, as suggested by the analysis of near-IR CMDs of the cluster (Chen et al., in prep.). Therefore the observational indication is consistent with our expectations at $\Zini =0.004$. 

\subsection{Comparing the C-rich LFs in the LMC and SMC}

\begin{figure*}
    \centering
    \includegraphics[width=0.7\textwidth]{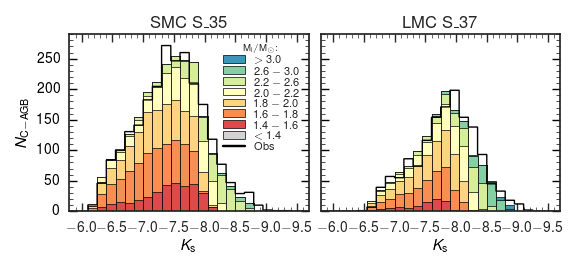}
    \includegraphics[width=0.7\textwidth]{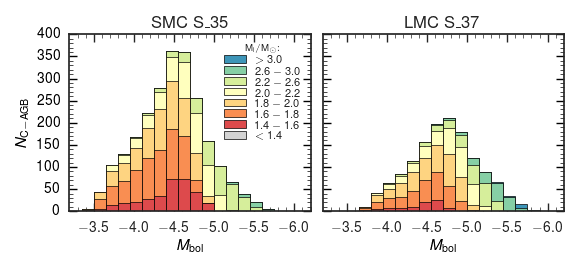}    
    \caption{Top panel: observed C-rich star luminosity functions in the absolute \ks\ magnitude (black solid histograms). Best-fitting models (set S\_35 for the SMC and set S\_37 for the LMC) are over-plotted  as stacked histograms binned as a function of the initial stellar mass. Bottom panel: theoretical C-rich star luminosity distributions in absolute bolometric magnitude.}
    \label{fig:cslf}
\end{figure*}

At this point it is interesting to compare the C-rich LFs in the Magellanic Clouds. The observed distributions in the absolute \ks\ magnitude are presented in the top panel of Fig.~\ref{fig:cslf}.  Distances and reddening corrections are taken from the results of the SFH recovery, described in Sect.~\ref{ssec:sfh} for the LMC and in \citet{pastorelli19} for the SMC.
We should recall here that the observed sample of carbon stars in the LMC is extracted from a selection of VMC tiles which do not cover the entire galaxy (see Sect.~\ref{ssec:sfh}). Anyhow, we can derive some useful indications coupling observations and best-fitting models.

In the absolute \ks\ magnitude the C-rich LFs present a similar morphology, but with two main evident differences.
First, in the LMC the peak is shifted towards brighter magnitudes (by $\approx$ 0.6 mag) compared with the SMC.
Similarly, a small magnitude shift in the $J$-band C-rich LF peaks of the two galaxies  has been recently reported and analysed by \citet[][see also Fig.~\ref{ssec:lfs_tmass_spitzer} for the LFs in 2MASS filters]{ripoche20}.
We interpret such a difference as due to the characteristics of the 3DU which is, on average, somewhat less efficient in the LMC, likely driven by a metallicity effect. At the same time the minimum mass for carbon star formation is expected to be larger at higher metallicity (see Sect.~\ref{ssec:cstars}). Second, the observed C-rich LF  extends to fainter \ks\ magnitudes in the SMC, since at lower metallicity the transition to the carbon star domain takes place earlier on the TP-AGB, that is at fainter magnitudes (see Fig.~\ref{fig:mbol_tr_lum}).

Similar considerations apply to the predicted C-rich LFs in bolometric magnitude (bottom panel of Fig.~\ref{fig:cslf}): in the SMC the peak appears located at $\Mbol \approx -4.45$, while the maximum of the LMC distribution  is at  $\Mbol \approx -4.75$.

\subsection{Mass-loss rates}
\label{ssec:mloss_rates}

\begin{figure}
    \centering
    \includegraphics[width=\columnwidth]{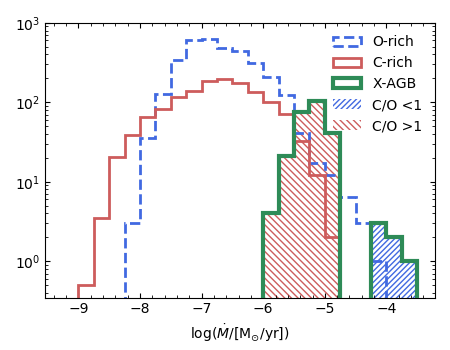}
    \caption{Predicted mass-loss rates distributions for the O-, C-, and X-AGB synthetic populations from the set S\_37. The X-AGB stars with C/O > 1 and those with C/O < 1 are shown as red and blue hatched regions, respectively, as indicated in the legend.}
    \label{fig:mloss_rates}
\end{figure}

The distributions of the mass-loss rates of the O-, C-, and X-AGB synthetic populations from the set S\_37 are shown in Fig.~\ref{fig:mloss_rates}. 
These results are similar to our previous SMC analysis, and are in broad agreement with the results of the SED fitting works by \citep{nanni19,groenewegen_sloan_18}. 
The majority of C- and O-rich stars have mass-loss rates in the range $\approx 10^{-8} - 10^{-6}~\MsunYr$. 
The X-AGB stars show the highest values of mass-loss rates, ranging from $\approx 10^{-6} - 10^{-4}~\MsunYr$. Specifically, the highest mass-loss rates are achieved by obscured O-rich AGBs experiencing HBB and with initial masses larger than $3~\Msun$. 
The percentage of O- and C-rich X-AGB stars in our simulations are $\approx 2$ and $\approx 98$ per cent respectively. 
This is in agreement with the general idea that the X-AGB sample is mainly populated by C-rich stars \citep{vanloon05, dellagli15}

\subsection{Predicted TP-AGB lifetimes} 
\label{ssec:lifetimes}

\begin{figure*}
\centering
\includegraphics[width=0.9\textwidth]{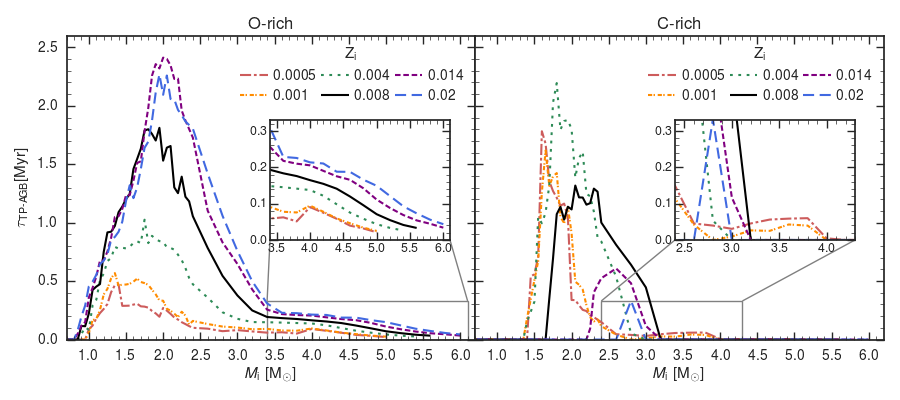}
\caption{TP-AGB lifetimes of O- and C-rich stars for selected values of initial metallicities as predicted by S\_35 for $\Zini =0.0005, 0.001, 0.004$ \citep[SMC calibration][]{pastorelli19} and by the S\_37 for $\Zini=0.008, 0.014, 0.02$ (LMC calibration, this work).
The lifetimes for the more massive stars are shown in the two insets of each panel.
As explained in the text, the lifetimes for $\Zini \geq 0.014$ are likely to be revised. }
\label{fig:lifetimes}
\end{figure*}

The reproduction of the observed star counts allows us to directly constrain the duration of the TP-AGB phase in the metallicity range considered. 
The lifetimes of TP-AGB stars are closely linked to their contribution to the integrated light of galaxies and to the total ejecta that they can return to the interstellar medium.

In Fig.~\ref{fig:lifetimes}, we show the lifetimes for O- and C-rich stars as a function of initial metallicity, which result from our SMC and LMC calibration. Specifically, we plot the predicted lifetimes extracted from the  \colibri\ set S\_35 for $\Zini =0.0005, 0.001, 0.004$ (SMC calibration) and from the S\_37 for $\Zini=0.008, 0.014, 0.02$ (LMC calibration). We caution that the results for solar-like metallicities, $\Zini \geq 0.014$ need to be further checked, and likely revised, with the help of observations that better probe this metallicity regime (e.g. AGB stars in M31, or the IFMR  in Galactic open clusters). This work is underway.

The O-rich lifetimes tend to increase with the initial metallicity. This is the combined result of the pre-dust mass-loss prescription, which is efficient in low-mass low-metallicity AGB stars, and the longer duration of the O-rich stages given that it is more difficult to form carbon stars at higher \Zini.
As to the C-rich lifetimes, a mirror-like trend is predicted. At larger \Zini, on average,  the duration is shortened since the TP-AGB phase is dominated by the O-rich stages. Note that the peak in the lifetimes of C-rich stars tends to shift to larger initial masses at increasing \Zini.

\subsection{TP-AGB contribution to integrated luminosity}
\label{ssec:int_mag}

\begin{figure*}
\centering    
\includegraphics[width=0.7\textwidth]{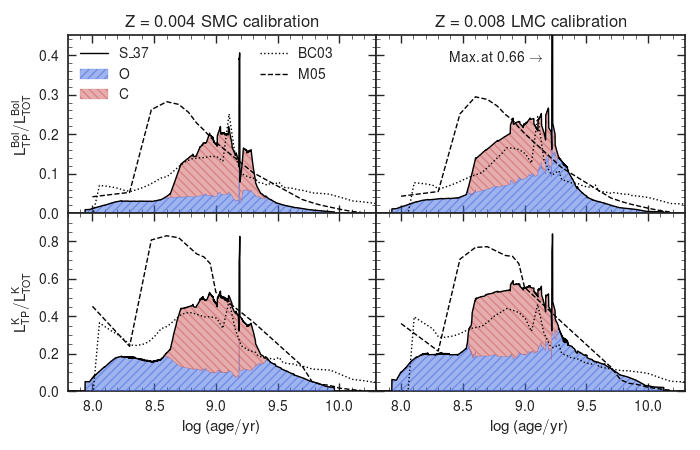}
\caption{Predicted contribution of TP-AGB stars to the total bolometric luminosity (upper panel) and the $K$-band luminosity (lower panel), as a function of age, from the \colibri\ set S\_37 at $\Zini = 0.004$ and $\Zini = 0.008$ (solid black lines).
The results for $\Zini = 0.004$ are based on the SMC calibration from \citealt{pastorelli19}, while the results for $\Zini = 0.008$ are from the LMC calibration discussed in this work. The red and blue hatched regions correspond to the contributions of C- and O-rich stars, respectively. We compare our results  with \citet[][M05]{Maraston_etal_05} and \citet[][BC03]{BC03} models, which are shown as black dashed lines and black dotted lines respectively.}
\label{fig:int_mags}
\end{figure*}

As one of the main motivations for providing calibrated TP-AGB evolutionary tracks and isochrones resides on the impact of this phase to the integrated light of galaxies, we compute the predicted contribution of TP-AGB stars to the luminosity of simple stellar populations (SSPs). 
 Fig.~\ref{fig:int_mags} shows this contribution, in both bolometric and $K$-band light, at $\Zini = 0.004$ and $\Zini = 0.008$. 
 The evolutionary tracks at $\Zini = 0.004$ and $\Zini = 0.008$ used to compute the TP-AGB contributions to the SSPs are calibrated in the SMC \citepalias{pastorelli19} and in the LMC (this work), respectively. 
 
We note that the spike at $\mathrm{\log(age/yr)\simeq 9.2}$ corresponds to the occurrence of the `AGB-boosting' effect, which is intimately linked to the physics of the stellar structure \citep[refer to][for a thorough discussion]{girardi13}.

In general, our predicted TP-AGB contribution peaks around $\sim1$~Gyr and does not exceed $\simeq$ 50-55 per cent in the $K$-band luminosity ($\simeq$ 20-25 per cent in bolometric luminosity). 
For ages younger than $\log(\mathrm{age/yr}) \approx 8.5$ and older than $\log(\mathrm{age/yr}) \approx 10$, the predicted TP-AGB contribution is almost entirely due to O-rich stars and increases with metallicity, as the O-rich lifetimes. 
The predicted contribution around the peak slightly increases from $\Zini = 0.004$ to $\Zini = 0.008$ (for both $K$-band and bolometric luminosity). 

We also compare our results to the popular stellar population synthesis models by \citet[][BC03]{BC03} and \citet[][M05]{Maraston_etal_05}. 
Our models predict a lower contribution with respect to M05 for which the peak in bolometric luminosity reaches 30 per cent, and it is as high as 80 per cent in the $K$-band at $\Zini = 0.004$. 
Furthermore, the peak in M05 is shifted to younger ages with respect to both our results and BC03. 
In this respect, we note that such differences are expected to be mitigated if we consider the work by \citet{noel_et_al_2013} which suggest a significant reduction of the TP-AGB contribution in M05 models based on a new analysis of the integrated colours of MCs' clusters. 
The results of our models are in general closer to BC03. The main differences in this case are the higher contribution predicted by BC03 for $\log(\mathrm{age/yr}) \lesssim 8.5-8.7$ and $\log(\mathrm{age/yr}) \gtrsim 9.5$ for both the bolometric and $K$-band luminosity.

\section{Conclusions}
\label{sec:conclu}

In this work we extend the calibration of \colibri\ TP-AGB models to the LMC.
To this purpose, we couple high-quality data of the AGB population in the LMC, with detailed stellar population synthesis simulations computed with the \trilegal\ code. 
We make use of the catalogue of AGB stars photometrically classified by \citet{boyer11}, and complemented with spectroscopic information from \citet{boyer15, groenewegen_sloan_18, kontizas01, MKtypes}. We revise the classification of a-AGB stars by using their location in the \textit{Gaia}-2MASS diagram, that allows a better separation between O- and C-rich stars with respect to the 2MASS and \textit{Spitzer} colours alone. 
The \trilegal\ simulations are based on the spatially resolved SFH derived from the deep near-infrared data of the VMC survey. Despite the partial coverage of the current SFH data, and the constraints on the quality of the SFH solutions (i.e. we consider only VMC subregions for which the RGB number counts are reproduced within $3\sigma$) the observed sample contains a statistically significant number of AGB stars ($\approx 4600$). 
We compare the observed and simulated \ks-band LFs of the O-, C-, X-AGB samples to investigate the 3DU at LMC-like metallicities. 

We reproduce the LMC photometry with the best-fitting set (S\_37), that has the same mass-loss prescription as the starting set (S\_35), but a reduced efficiency of the 3DU for metallicities higher than $\Zini = 0.008$.
Such a decreasing trend of the 3DU efficiency is in line with the results of the full TP-AGB models by \citet{Cristallo_etal_11} and \citet{VenturaDantona_09}.

As new molecular linelists for species relevant to the C-rich star opacity recently became available, a new set of \textsc{comarcs} models is currently being calculated. In this work we test how the new opacity data affect the predicted colours of C-rich stars. We find that a significant improvement in the LMC \jks\ colour distribution and \ks-LFs can be achieved with the new models. However, the new grid of \textsc{comarcs} models is not yet complete and cannot be consistently used in \trilegal\ for the calculation of synthetic photometry. We also show that the new data will not affect our previous SMC calibration. 

We also find a fairly good agreement between the predicted and observed LFs in the 2MASS and \textit{Spitzer} filters.

We then discuss our best-fitting model in terms of carbon star formation, mass-loss rate distributions, stellar lifetimes and predicted contribution to the integrated light of SSPs. 
The C-rich star domain at $\Zini = 0.008$ extends from $\Mini \approx 1.7~\Msun$ to $\Mini \approx 3~\Msun$, whereas at $\Zini = 0.004$ from $\Mini \approx 1.4~\Msun$ to $\Mini \approx 2.8~\Msun$. The minimum initial mass for C-rich star formation is predicted to decrease with decreasing metallicity, in agreement with other works in the literature. 
Our results are in agreement with the mass ranges inferred from LMC clusters. The mass-loss rate distributions predicted by our best-fitting model are in broad agreement with the results of the SED studies by \citet{groenewegen_sloan_18} and \citet{nanni19}.
With respect to the SMC calibration (S\_35), the best-fitting model S\_37 predicts longer lifetimes for O-rich stars and shorter lifetimes for C-rich stars at $\Zini = 0.008$.
The peak of the TP-AGB contribution to the bolometric and $K$-band luminosity predicted by our best-fitting model S\_37 at $\Zini = 0.008$ and ages around 1 Gyr lies in between the predictions by BC03 and M05. 

New sets of \parsec+\colibri\ stellar isochrones derived from the best-fitting model presented in this work (S\_37) are available via our web interface \url{http://stev.oapd.inaf.it/cmd}. We refer to \citet{marigo17} for all the details concerning the  construction and the format of the provided isochrones. 
The current stellar isochrones also include the predicted pulsation periods of long-period variables from  \citet{trabucchi19}. The new set of bolometric corrections by \citet{YBCpaper} are also available, and will become the default option for subsequent works.
The present calibration of \colibri\ TP-AGB models covers the sub-solar metallicity range relevant for the Magellanic Clouds ($\Zini = 0.0005 - 0.01$). We provide isochrones also 
for higher metallicities, i.e. $\Zini=0.014, 0.017, 0.02$, which cover the solar-like metallicity range. In this respect we caution the users that these models may need to be revised with the aid of observational constraints that suitably  probe the solar-metallicity regime. Work is in progress using AGB stars in M31 and the IFMR in the Milky Way. The same applies to very low metallicities using data for Local Group dwarf galaxies.
Further studies are also planned for the LMC, once the new grid of \textsc{comarcs} model spectra for AGB stars and the SFH of more VMC tiles become available. In particular, we will explore in more details the dependence of our calibration on the age (hence initial mass) of AGB stars by studying the LFs of regions in the Magellanic Clouds characterized by young and old stellar populations separately.


\section*{Acknowledgements}
Many of us (PM, GP, LG, SB, YC, BA, MT, MATG, JD, SR) acknowledge the support from the  ERC Consolidator Grant funding scheme ({\em project STARKEY}, grant agreement n. 615604).
We thank C. Maraston, S. Charlot, and G. Bruzual for providing us with their stellar population synthesis models. 
M-RC acknowledges support the European Research Council (ERC) under the European Union’s Horizon 2020 research and innovation programme (grant agreement no. 682115).
AN acknowledges the support by the Centre National d'Etudes Spatiales (CNES) through post-doctoral fellowship.
This work is based on the observations collected at the European Organisation for Astronomical Research in the Southern Hemisphere under ESO programme 179.B-2003. We thank the CASU and the WFAU for providing calibrated data products under the support of the Science and Technology Facility Council (STFC) in the UK.
This publication makes use of data products from the Two Micron All Sky Survey, which is a joint project of the University of Massachusetts and the Infrared Processing and Analysis Center/California Institute of Technology, funded by the National Aeronautics and Space Administration and the National Science Foundation.
This work has made use of data from the European Space Agency (ESA) mission {\it Gaia} (\url{https://www.cosmos.esa.int/gaia}), processed by the {\it Gaia} Data Processing and Analysis Consortium (DPAC, \url{https://www.cosmos.esa.int/web/gaia/dpac/consortium}). Funding for the DPAC has been provided by national institutions, in particular the institutions participating in the {\it Gaia} Multilateral Agreement.
This research made use of Astropy, a community-developed core Python package for Astronomy \citep{astropy13, astropy18} and matplotlib, a Python library for publication quality graphics \citep{matplotlib}.

\section*{Data Availability}
The stellar models and isochrones generated in this research will be shared via the CMD web interface at \url{http://stev.oapd.inaf.it/cmd}.
The VMC data used in this research will be soon shared by ESO via its regular data releases (see \url{http://www.eso.org/rm/publicAccess#/dataReleases}).
The catalogues of observed AGB stars underlying this article are available from \citet{boyer11} at 10.26093/cds/vizier.51420103, 
\citet{groenewegen_sloan_18} at 10.26093/cds/vizier.36090114,
\citet{kontizas01} at 10.26093/cds/vizier.33690932, 
\citet{MKtypes} at \url{http://vizier.u-strasbg.fr/viz-bin/VizieR?-source=B/mk}.
The spectroscopic classification data presented in \citet{boyer15} and used in this research were provided by M. L. Boyer by permission. Data will be shared on reasonable request to the corresponding author with permission of M. L. Boyer.


%
\input{main.bbl}
\label{lastpage}
\end{document}